\documentclass[aps,prb,twocolumn,superscriptaddress]{revtex4-2}

\usepackage{amsmath, amssymb,  graphicx}

\usepackage[colorlinks=true ,urlcolor=blue,urlbordercolor={0 1 1}]{hyperref}

\usepackage{color}
\usepackage{xcolor}
\usepackage{braket}

\usepackage[utf8]{inputenc}
\usepackage{bm}
\usepackage{booktabs}
\usepackage{verbatim}
\usepackage{graphicx}
\usepackage{amsmath}
\usepackage{color}
\usepackage{float}
\usepackage{bbm}
\usepackage{amssymb}
\usepackage{slashed}
\usepackage{wasysym,bm,bbm,dsfont}
\usepackage{makecell}

\begin{document}

\title{Mixed valence Mott insulator and composite excitation in twisted bilayer graphene}
\author{Jing-Yu Zhao}
\author{Boran Zhou}
\author{Ya-Hui Zhang}
\affiliation{Department of Physics and Astronomy, Johns Hopkins University, Baltimore, Maryland 21218, USA}

\date{\today}

\begin{abstract}
Interplay of strong correlation and flat topological band has been a central problem in moir\'e systems such as the magic angle twisted bilayer graphene (TBG). Recent studies show that Mott-like states may still be possible in TBG despite the Wannier obstruction. However, the nature of such unconventional states is still not well understood. In this work we construct the ground state wavefunction and exotic excitations of a symmetric correlated semimetal or insulator at even integer filling using a parton mean field theory of the topological heavy fermion model (THFM). We label the valence of the $f$ orbital in the AA site based on its occupation $n_f$.   At $\nu=-2$, we show that the $f$ orbital is not in the simple $f^{2+}$ valence expected from a trivial Mott localization picture. Instead, around $1/3$ of AA sites are self doped,  with holes entering the $c$ orbitals away from AA sites.  As a result, the $f$ orbital is in a superposition of $f^{2+}$ and $f^{3+}$ valences and should not be viewed as local moment.  Based on this picture, we dub the phase as \textit{mixed valence Mott insulator}. This unconventional insulator still has a large hybridization $\langle c^\dagger f \rangle\neq 0$  and is sharply distinct from the usual `kondo breakdown' picture. In most of the momentum space away from the $\Gamma$ point, there is  a Mott gap equal to the Hubbard $U$. On the other hand, at the $\Gamma$ point, we only have a `charge transfer gap' much smaller than $U$. In particular, the top of the lower band is dominated by a composite excitation, which is a linear combination of $\ket{f^{1+}}\bra{f^{2+}}$ and $\ket{f^{2+}}\bra{f^{3+}}$ with a sign structure such that it is orthogonal to the microscopic $f$ operator. At $\nu=0$, similar approach leads to the description of a Mott semimetal.  We hope this work will inspire more explorations of the Anderson models with a large hybridization, a regime which  may host new physics beyond the familiar Kondo or heavy fermion systems.
\end{abstract}

\maketitle

\textbf{Introduction} Understanding the nature of correlated insulating states and superconductivity in magic-angle twisted bilayer graphene (TBG) \cite{cao2018unconventional,Cao2018mott,Yankowitz2019layer,Lu2019moreSC,Stepanov2020,Cao2021nematic,Liu2021Coulomb,arora2020superconductivity} remains one of the central open problems in the study of moiré materials. 
In particular, the insulating state at filling $\nu = -2$ and the superconducting dome emerging upon doping ($\nu = -2 - x$) have attracted significant attention due to their intriguing similarity to the phase diagram of high-temperature cuprates. 
However, due to a non-trivial fragile topology of the active flat band
\cite{Po2018IVC,Tarnopolsky2019,Ahn2019fgtop,Ledwith2021Peda,Song2021symano}, 
a simple lattice model is impossible and the correlated insulator must be different from a conventional Mott insulator.  Instead, various isospin polarized phases\cite{bultinck2020ground,kwan2021kekule,parker2021strain,wagner2022global} were found from Hartree-Fock calculations. On the other hand, experimental measurements of entropy indicate the existence of local moments at finite temperature\cite{rozen2021entropic,saito2021isospin}. Therefore some kind of Mott physics may still be relevant when the generalized ferromagnetic state is thermally disordered. It is an important theoretical challenge to develop frameworks to capture such `Mott' states and elucidate the distinction from the conventional Mott insulator in trivial band.

There have already been progresses in formulating real space models through the topological heavy fermion model\cite{Song2022THFM,Calugaru2023THFM,Yu2023THFM,Hu2023THFM,Vafek2024THFM,Zhou2024THFM,Hu2023THFM2,Chou2023THFM,Wang2024THFM,Lau2023THFM,Rai2024THFM,Youn2024DMFT,checkelsky2024flat} or a model with non-local orbital\cite{Ledwith2024}. Recently we also proposed a new theoretical framework to understand potential topological Mott states\cite{Zhao2025ancTBG} directly in the momentum space through the ancilla fermion technique\cite{Zhang2020,Zhou2024}. There are now indications that there exist  Mott-like states at integer fillings\cite{Ledwith2024,Hu2025THFM,Zhao2025ancTBG,ledwith2025exotic}, which have well separated Mott gap in momentum space away from the $\Gamma$ point. Around the $\Gamma$ point, there is a quadratic band crossing at $\nu=0$ and a small gap at other integer fillings. Such a momentum selective Mott gap can be simply understood in the ancilla theory\cite{Zhao2025ancTBG}. In this framework, the Mott gap emerges from an exciton-like hybridization $\Phi(\mathbf{k})c^\dagger_{\mathbf{k}}\psi_{\mathbf{k}}^{}$ between the physical electron $c$ and an ancilla fermion $\psi$. In trivial band $\Phi(\mathbf k)=\frac{U}{2}$ is a constant in momentum space. In contrast, the nontrivial band topology of TBG leads to a nodal order parameter with  $\Phi(\mathbf k)\sim k_x \pm i k_y$ around $\mathbf k=0$. As a result, the gap at $\mathbf k=0$ vanishes for $\nu=0$. At non-zero integer filling, there is a small gap from the relative energy shift between $\psi$ and $c$.

Although these previous works successfully calculated the spectrum of the Hubbard bands, the nature of the Mott-like states is still poorly understood. For example, what is the charge distribution of these states? In a conventional Mott insulator at large Hubbard U, we just have a fixed integer number of electrons at each site. In TBG, such a perfect Mott localization is obstructed. To answer this question, we need to understand the ground state wavefunction of these states, which is beyond the Green's function approach in Ref.\cite{Ledwith2024,Hu2025THFM}. Wavefunctions were constructed in the ancilla theory\cite{Zhao2025ancTBG}, but  there lacks a clear physical interpretation in terms of microscopic degrees of freedom. It is clear that a new theoretical approach is needed to fully reveal the nature of these unconventional states.

In this work we describe symmetric correlated states at even integer fillings using a more conventional parton theory, as a generalization of the familiar slave boson or Schwinger boson theory. To formulate the theory, we use the topological heavy fermion model (THFM) \cite{Song2022THFM}. However we  emphasize that we use the model simply for convenience and the physics we discover is very different from the familiar Kondo or heavy Fermion systems. In THFM there are dispersive $c$ bands  hybridizing with a flat wannierized $f$ orbital.   The $f$ orbital is well localized on the AA site and we label its valence based on the occupation $n_f$. For example, $f^{2+}$ indicates a state with two electrons in the $f$ orbital.  Without the $c$ orbital, at $\nu=-2$, we just have a simple Mott insulator with $f$ orbital always in the $f^{2+}$ valence. Our key discovery is that in THFM, the $f$ orbital is self doped by creating holes in the $c$ orbital. As a result, the $f$ orbital is in a superposition of $f^{2+}$ and $f^{3+}$ state and has a valence around $f^{2.3+}$ on average.  Remarkably, due to the large hybridization, the state still has a finite insulating gap and feature unconventional excitation around $\mathbf k=0$. The new approach enables us to explicitly identify the microscopic content of the composite excitation and match it with the ancilla fermion in our previous theory\cite{Zhao2025ancTBG}. Especially now we have a deeper understanding why such an exotic excitation dominates at low energy. 

Based on the new understanding, we dub the phase as \textit{mixed valence Mott insulator}. The state is at $1/2$ filling per spin-valley flavor and is different from the familiar mixed-valence insulator which is at integer filling and connects to a band insulator.  In the conventional slave boson treatment\cite{coleman2015introduction,coleman1984new} of the Anderson model, one usually finds a heavy Fermi liquid phase or a Kondo breakdown phase based on whether the hybridization $\langle c^\dagger f \rangle$ vanishes or not. In contrast, here we find a state which connects to the Mott insulator, but still has large $\langle c^\dagger f \rangle$ (see Table~\ref{tab:mixedvalence}). To our best knowledge, such a possibility has not been well explored in the previous theoretical treatments of the Anderson model or heavy Fermion systems.

\begin{table}[t]
    \centering
    \begin{tabular}{l|c|c}
        \hline
          &  $\langle c^\dagger f\rangle\neq 0$ &  $\langle c^\dagger f\rangle=0$ \\
        \hline
        $A_{\mathrm{FL}} - \frac{n}{4} =0 $& Heavy Fermi Liquid& $*$\\  \hline
        $A_{\mathrm{FL}} - \frac{n}{4} =1/2$ & \textbf{Mixed valence Mott }& Kondo breakdown\\
        \bottomrule
    \end{tabular}
    \caption{Different phases of Anderson Model around $\nu=-2$ or $n=4+\nu=2$. $\frac{n}{4}$ is the filling per spin-valley flavor. $A_{\mathrm{FL}}$ is the Fermi surface volume (mod 1) per spin-valley flavor. At $n=2$, $A_{\mathrm{FL}}-\frac{n}{4}=\frac{1}{2}$ means $A_{\mathrm{FL}}=0$ mod 1, indicating an insulator.}
    \label{tab:mixedvalence}
\end{table}

\begin{figure}[ht]
    \centering
    \includegraphics[width=0.95\linewidth]{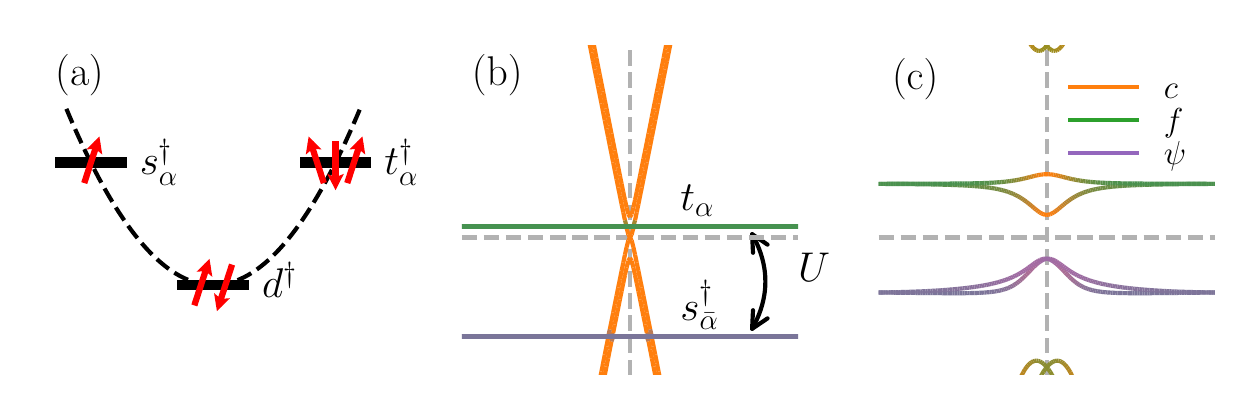}
    \caption{Illustration of our mean field description of the $\nu=-2$ insulator. 
    (a) For each orbital $f_i$, we consider the $f^{1+}$, $f^{2+}$, $f^{3+}$ valences denoted as $s^\dagger_\alpha\ket{0}, d^\dagger\ket{0}$ and $t^\dagger_\alpha\ket{0}$. For the doublon $d$, we restrict to  the  spin-singlet state due to anti-Hund's coupling. 
    (b) We illustrate a simple   Kondo breakdown phase where the $f$ orbital is decoupled with the conductive $c$ bands. The ground state of $f$ is a product of $\ket{d}$. The upper and lower hubbard bands are created by $t^\dagger_\alpha$ and $s_\alpha$ respectively. The system is in a semi-metal phase due to the gapless $c$ band.
    (c) On the other hand, with finite hybridization $\gamma\neq 0$, 
    the system is in an insulator, despite that $f$ orbital is self doped with a superposition of $f^{2+}$ and $f^{3+}$ valences.
    While $f\sim s+t^\dagger$ hybridizes with $c_1$ and is pushed to the remote band, $\psi\sim -s+t^\dagger$ remains as a cheap excitation at $\mathbf k=0$ upon hole doping. }
    \label{fig:illu}
\end{figure}

\textbf{Model} We consider the THFM  description of TBG\cite{Song2022THFM},
\begin{equation}\label{eqn:THFM_0}
    H_0= H^{(c_1,c_2)}_0  + H_0^{(c_1, f)} + H_{\mathrm{int}}^{(f)}-\mu (N_c+N_f)~,
\end{equation}
which includes two dispersive bands, $c_1$ and $c_2$, characterized by a large bandwidth $v_*|\mathbf{K}|$, 
and a Wannier flat band $f$.
When a finite hybridization 
$\gamma(\mathbf{k}) = e^{-k^2\lambda^2/2}\left(\gamma\sigma_0 +v^\prime_\star(k_x\sigma_x+k_y\tau_z\sigma_y)\right)$ 
is introduced between $c_1$ and $f$, with 
\begin{equation}
    H_0^{(c_1,f)} = \sum_{\mathbf{k},\mathbf G} f_{\mathbf k}^\dagger \gamma(\mathbf k+\mathbf G) c_{1,\mathbf{k}+\mathbf G} + \mathrm{h.c.}~, 
\end{equation}
Eq.~\eqref{eqn:THFM_0} reproduces the band structure of the microscopic Bistritzer–MacDonald (BM) model~\cite{Bistritzer2011}.
Here, both $f_{\mathbf{k}}$ and $c_{1,\mathbf{k}}$ are eight-component spinors, collecting spin, valley, and orbital flavor: $f_{\mathbf{k}} = \{f_{\mathbf{k};\alpha}\}$ and $c_{1,\mathbf{k}} = \{c_{1,\mathbf{k};\alpha}\}$, with $\alpha = 1, \dots, 8$ labeling the flavor. $\alpha=a \tau s$ is formed by the orbital $a=\pm$, valley $\tau=K, K'$ and spin $s=\uparrow, \downarrow$. $f_{a=\pm;\tau s}$ has angular momentum $L=\pm 1$ around each AA site.

For the $f$ orbital,  we include simple on-site interaction;
\begin{equation}\label{eqn:interactionf}
    H_{\mathrm{int}}^{(f)} = U/2\sum_i(n_{i;f} - 4-\kappa\nu)^2 + H_{J_A}^{(f)}, 
\end{equation}
where we also include the anti-Hund's spin interaction term $H_{J_A}^{(f)}$ arising from electron–phonon coupling~\cite{Wang2024THFM}. 
$H_{J_A}^{(f)}$ favors inter-valley spin-singlet  as discussed in the Supplementary Material. We also follow Ref.~\cite{Lau2023THFM} and add a phenomenological attractive potential through $-\kappa \nu$. 
Its role is to shift the energy of $f$ orbital by $\delta E_f=-\kappa U \nu$ relative to the $c$ orbital. 
Such an energy shift may arise from repulsion between $c$ and $f$ orbital. We leave a microscopic derivation of this term to future work and treat it as a tunable parameter here. In agreement with Ref.~\cite{Lau2023THFM},  we find $\kappa\approx 0.8$ gives reasonable results, in the sense that the new approach macthes the previous ancilla theory\cite{Zhao2025ancTBG} in the spectrum of single electron excitation. 

\textbf{Effective model with restricted Hilbert space} We are interested in a symmetric correlated insulator at $\nu=-2$. So we need to add a relatively large anti-Hund's coupling $J_A$ to favor spin-singlet state instead of ferromagnetism. Unlike Ref.~\cite{Ledwith2024} focusing on finite temperature, we hope to obtain a ground state wavefunction.   If we ignore the $\gamma$ hybridization, it is easy to imagine a `Kondo breakdown phase' with $f$ orbital in the $f^{2+}$ valence, decoupled from a semimetal of $c$ bands.  Next we need to incorporate the hybridization $\gamma$. For the $f$ orbital, we can keep only $f^{1+},f^{2+},f^{3+}$ valences. In the following we label these states as singlon, doublon and triplon.  To make analytical treatment possible, we consider a simpler effective model with a restricted Hilbert space. We only keep the spin-singlet doublon state with angular moemntum $L=0$ (the ground state of the $J_A$ term at each site): $|d_i\rangle = \sum_{\alpha} \frac{\mathrm{sgn}(\alpha)}{2}f^\dagger_{i;\alpha}f^\dagger_{i;\bar\alpha}|0\rangle$, 
$\mathrm{sgn}(\alpha) = \pm 1$ distinguishes spin-up and spin-down flavors, and $\bar\alpha$ denotes the time-reversed partner of $\alpha$.   In this language, the $f$ orbital is in a simple product state $\prod_i \ket{d_i}$ in the decoupling limit.  Viewing $\ket{d}$ as a vacuum state, a finite hybridization  $\gamma f^\dagger c_1+h.c.$ excites a subspace of the $f^{1+}$ and $f^{3+}$ valences, 
which include singlon state $|s_{i;\bar\alpha}\rangle =2f_{i;\alpha}|d_i\rangle$; 
and triplon state 
$|t_{i;\alpha}\rangle=\frac{2}{\sqrt{3}}f^\dagger_{i;\alpha}|d_i\rangle$. The normalization factor is chosen to make $\braket{s|s}=1$ and $\braket{t|t}=1$.    We now introduce two new fermionic operators: $ s^\dagger_{i;\alpha} = |s_{i;\alpha}\rangle \langle d_i|$,
$ t^\dagger_{i;\alpha} = |t_{i;\alpha}\rangle \langle d_i|$. It is easy to see that $s^\dagger, t^\dagger$ have the same spin representation as the $f^\dagger$ operator.  We also define $n_{i;s}=\sum_{\alpha} \ket{s_{i;\alpha}}\bra{s_{i;\alpha}}$ and similarly $n_{i;t}$ and $n_{i;d}$.

In the effective model we only include $\ket{s_\alpha}, \ket{t_\alpha}, \ket{d}$ state at each AA site. We do not impose any constraint on the $c$ orbital.  We must have $n_{i;s}+n_{i;t}+n_{i;d}=1$, so $n_{i;s}+n_{i;t}\leq 1$.  We also have the physical $f$ density $n_{i;f}=n_{i;t}-n_{i;s}+2$.  The constraint $n_{i;s}+n_{i;t} \leq 1$ is similar to the Gutzwiller constraint in the familiar $t-J$ model.  We label the projection operator as $P_G$, which restrict the Hilbert space of each $f$ site to include only the above states. Note here $\ket{d}$ should be viewed as a vacuum state, similar to the $\ket{0}$ state in the $t-J$ model.

After the projection, the physical $f$-electron operator can be expressed as
\begin{equation}
    P_G f^\dagger_{i;\alpha} P_G = \frac{1}{2} s_{i;\bar\alpha} + \frac{\sqrt{3}}{2} t^\dagger_{i;\alpha},
    \label{eq:f_operator}
\end{equation}

We can project the full Hamiltonian Eq.~\eqref{eqn:THFM_0} into this reduced Hilbert space, $H_0 \rightarrow P_G H_0 P_G$ by replacing $f$ with $P_G f P_G$ and 
the interaction Hamiltonian with 
\begin{equation}\label{eqn:THFM_U_esd}
    P_G(H_{\mathrm{int}}^{(f)} -\mu N_f)P_G = \sum_i(E_s n_{i;s} + E_t n_{i;t}) + \mathrm{const.}\\
\end{equation}
Where  $E_s = U/2+2U(1-\kappa)+\mu$ and $E_t=U/2-2U(1-\kappa)-\mu$ are on-site energy of singlon and triplon excitations.

Unlike the usual slave boson theory of the $t-J$ model, here we actually have two fermionic operators $s$ and $t$.  One specific linear combination in Eq.~\ref{eq:f_operator} gives the microscopic $f$ operator within the subspace.  It is also useful to define another fermionic operator from an orthogonal linear combination:

\begin{equation}
    \psi_{i;\alpha}^\dagger  = -\frac{\sqrt{3}}{2}  s_{i;\bar\alpha}^{} + \frac{1}{2} t^\dagger_{i;\alpha}. \label{eq:psi_operator}
\end{equation}

Microscopically $\psi_{\alpha}$ corresponds to a composite operator: $\psi^\dagger_{i;\alpha} = P_G \big(A\{f^\dagger_{i;\alpha}, n_{i;f}\} - Bf^\dagger_{i;\alpha} \big) P_G$ with $A=2/\sqrt{3}, B=3\sqrt{3}$.  It corresponds to a microscopic trion operator\cite{Zhao2025ancTBG,ledwith2025exotic}, but it is easier to just view it as a different linear combination of $s$ and $t^\dagger$ operator within our subspace.

\begin{figure}[t]
    \centering
    \includegraphics[width=0.98\linewidth]{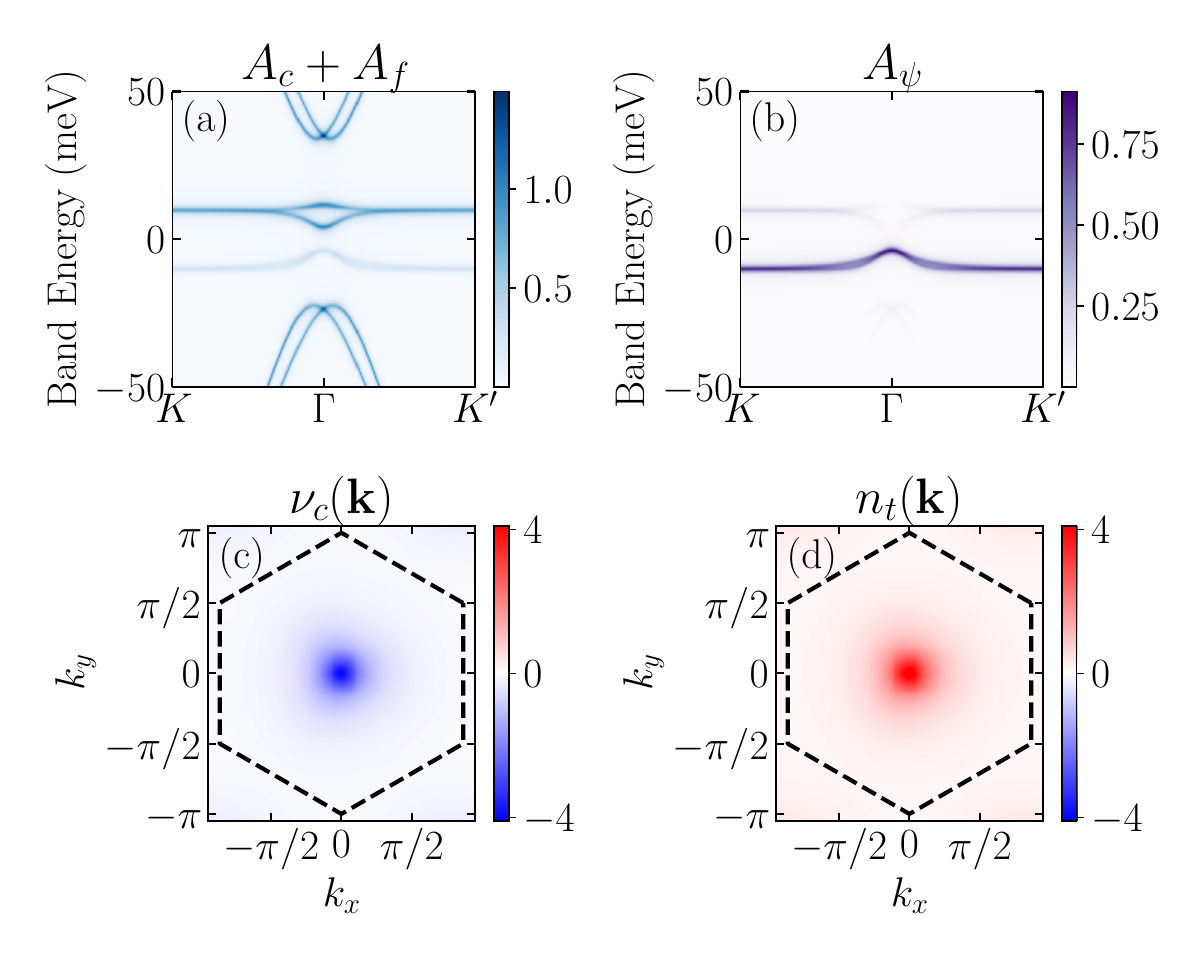}
    \caption{Typical mean field band structure at $\nu=-2$ using $U=20$meV, $\gamma=-38$meV, $\kappa=0.8$. (a) The total physical electron spectrum. 
    (b) Spectrum of the auxiliary fermion $\psi$, which is not directly observable in experiments. 
    (c) and (d) shows the momentum space density distribution of $\nu_c(\mathbf{k})=n_c(\mathbf k)$ and $n_t(\mathbf{k})$. One can see extra triplon excitations in $f$ orbital and hole excitations in $c$ orbital around $\mathbf k=0$, suggesting a mixed-valence nature of $f$ orbital around $\mathbf k=0$. In contrast, the state in momentum space away from $\mathbf k=0$ is in the fixed $f^{2+}$ valence, as in a conventional Mott insulator.
    }
    \label{fig:spectrum}
\end{figure}

\textbf{Renormalized mean field theory} We can perform a  mean field calculation by relaxing the Gutzwiller constraint  $P_G$, with a cost of including a renormalization factor \cite{zhang_rmft_1988, vollhardt_rmft_1984} for the $\gamma$ term. We  use the following mean field Hamiltonian without constraint: 
\begin{equation}\label{eqn:THFM_mf}
\begin{aligned}
    H_{\mathrm{MF}} =& H_0^{(c_1,c_2)}  + \sum_i(E_s n_{i;s} + E_t n_{i;t}) -\mu N_c + \\
    & {g_\gamma} \sum_{\mathbf k,\mathbf G\alpha,\beta} 
    (\frac{1}{2}s_{-\mathbf k}+\frac{\sqrt{3}}{2}t_{\mathbf k}^\dagger) \gamma(\mathbf{k+G})c^{}_{1,\mathbf{k+G}}+\mathrm{h.c}, 
\end{aligned}
\end{equation}
where $s_{\mathbf{-k}}$ and $t_{\mathbf{k}}$ are eight-component spinors $s_{-\mathbf{k}} = \{s_{-\mathbf{k};\bar\alpha}\}$, $t^\dagger_{\mathbf{k}} = \{t^\dagger_{\mathbf{k};\alpha}\}$. 

Now we treat $s$ and $t$ as free fermions,  
and normalize the hybridization term $\langle f^\dagger c\rangle$ with a factor $g_\gamma = \sqrt{1-\langle n_s\rangle -\langle n_t\rangle}$ due to the restriction $n_{i;s}+n_{i;t}\leq 1$. 
It is equivalent to a slave boson theory with $\ket{d_i}=d^\dagger_i \ket{0}$ and then consider slave boson condensation $\langle d \rangle=\sqrt{n_d}=\sqrt{1-n_s-n_t}$.  The ground state wavefunction is simply $\ket{\Psi_c}=P_G \ket{\textrm{Slater}}$, where the Slater determinant is the ground state wavefunction of the above mean field Hamiltonian.

\begin{figure}
    \centering
    \includegraphics[width=0.95\linewidth]{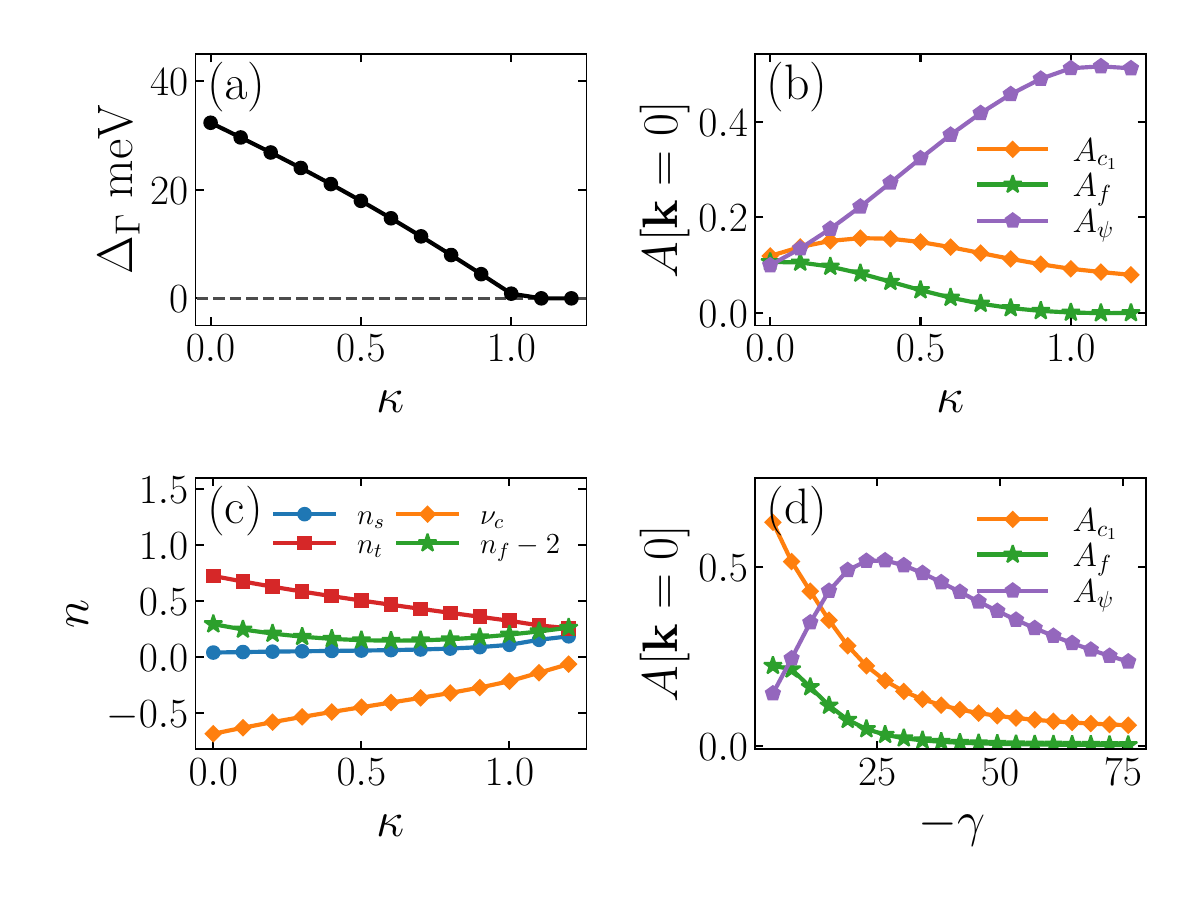}
    \caption{
    (a) $\Gamma$ point gap $\Delta_\Gamma$ as a function of $\kappa$ with fixed interaction $U=20$ meV and hybridization $\gamma=-38$ meV. There is an insulator to semimetal transition at $\kappa_c \approx 1$.  
    (b) The $\Gamma$-point spectral weight of $c_1$, $f$, and $\psi$ on the hole-doped side as a function of $\kappa$.
    $A_{c_1}$ increases slightly when $\kappa \ll 1$, but both $A_{c_1}$ and $A_f$ decrease at large $\kappa$.
    In contrast, $A_\psi$ increases and becomes dominant at large $\kappa$.
    (c) shows density $n_s$, $n_t$, $\nu_c$, $n_f$ as a function of $\kappa$. 
    (d) The $\Gamma$-point spectral weight of $c_1$, $f$, and $\psi$ on the hole-doped side as a function of $\gamma$ for $U = 20$ meV and $\kappa=0.8$. 
    Both $A_{c_1}$ and $A_f$ decrease with  $\gamma$.
    $A_\psi$ first increases with $\gamma$ and then start to decrease due to the decrease of $g_\gamma$ at large $\gamma$.
    }
    \label{fig:kappa}
\end{figure}

Typical band structures obtained from this mean-field analysis at $\nu=-2$ are shown in Fig.~\ref{fig:spectrum} for $U = 20$ meV and $\gamma=-38$meV. 
The spectrum $A_f$ and $A_\psi$ are multiplied by a factor $\langle n_d \rangle=(1-\langle n_s \rangle-\langle n_t \rangle)$ due to the restriction $n_{i;s}+n_{i;t}\leq 1$. 
Our results  are
in agreement with ancilla theory presented in Ref.~\cite{Zhao2025ancTBG}. Away from $\Gamma$ point, we can see the Mott gap at order of $U$. Close to the $\Gamma$ point, the gap is smaller and separate two different excitation. The excitation in the $E>0$ side is dominated by $c_2$, while the excitation in the $E<0$ side is dominated by $\psi$, which has vanishing spectral weight in the usual ARPES measurement. 

We can now have a deeper understanding of the nature of the  state here given that we have the microscopic wavefunction. First, the $f$ orbital is not in the fixed $f^{2+}$ valence anymore. It turns out that the $f$ orbital is self doped by creating holes in  the $c$ orbitals. As a result, there is a superposition of the $f^{2+}$ and $f^{3+}$ valence. As shown in Fig.~\ref{fig:spectrum}(c)(d), there are extra holes $\langle n_c(\mathbf k) \rangle <0$ and triplon $\langle n_t(\mathbf k) \rangle >0$ in a small region around $\mathbf k=0$. Due to an extra factor of $8$, the total number of triplon excitation $n_t\sim 0.3-0.4$  is not negligible. Hence the $f$ orbital is in a mixed-valence regime and can not be simply viewed as a local moment. Naively, with extral electrons in the $f$ orbital, the system should be in a metallic phase. However, here due to the large hybridization $\gamma$, the system stays in an insulator, but with very unique excitations.  Away from $\mathbf k=0$, we have $n_c(\mathbf k)\approx 0, n_t(\mathbf k)=0, n_s(\mathbf k)=0$, so the state is formed by the doublon state $\ket{d}$. Then the single electron excitation is simply to add or remove one electron in the $f$ orbital. For example, the upper Hubbard band is created by $t^\dagger(\mathbf k)$ and the lower band is created by $s^\dagger(\mathbf k)$. They are separated by the Hubbard $U$, just like the trivial Mott insulator. 

In contrast, non-trivial excitations arise from the mixed-valence nature around $\mathbf k=0$. Now  the system has holes from the $c$ band and triplon excitations in this momentum space region.  Then if we create an electron, it is energetically favorable to just enter the $c$ band and annihilate the holes of $c$ which already exist in the ground state. On the other hand, the hole doped side is dominated by a quite exotic cheap excitation. Because the ground state is in a mixture of $f^{2+}$ and $f^{3+}$ valence around $\mathbf k=0$, one can remove an electron through either $\ket{f^{1+}}\bra{f^{2+}}\sim s^\dagger$ or $\ket{f^{2+}}\bra{f^{3+}}\sim t$. With $\kappa$ close to $1$, the energy costs of $f^{1+}, f^{2+}, f^{3+}$ are roughly $U/2, 0, U/2$. Thus the above two process have energy cost of $U/2$ and $-U/2$ respectively.  Therefore a linear combination $s^\dagger \pm t$ actually cost an energy smaller than $U$.  Among the two independent linear combinations, one of them is the $f$ operator. This excitation is energetically penalized because the $f$ state is pushed to the remote band by the large $\gamma$ term at $\mathbf k=0$. On the other hand, the other operator $\psi(\mathbf k)$ can avoid the energy cost of both $U$ and $\gamma$ and remains at low energy.  We can now clearly see that the non-trivial excitation $\psi(\mathbf k)$ is closely associated with the mixed-valence nature of the ground state.

In Fig.~\ref{fig:kappa} we also demonstrate the dependence on the parameter $\kappa$ and $\gamma$. When $\kappa>1$, the gap actually closes and we have a semimetal at $\nu=-2$, which was also obtained in the ancilla theory in the small $U$ regime.

\textbf{Equivalence to the ancilla theory} We now demonstrate that the renormalized mean-field theory introduced above for $\nu=-2$ is equivalent to the ancilla theory previously formulated as a variational wavefunction approach~\cite{Zhao2025ancTBG}. The mean field Hamiltonian for the charge sector in the ancilla theory takes the form:
\begin{equation}\label{eq:ancilla_mf} \begin{split}
    H_\mathrm{ancilla}=&H_0^{(c_1,c_2)}+H_0^{(c_1,f)}+\Phi\sum_\mathbf{k}\left(f^\dagger_{\mathbf{k}}\psi_{\mathbf{k}}+\mathrm{H.c.}\right)\\
    &-\mu N_c-\mu_fN_f-\mu_\psi N_\psi,
    \end{split}
\end{equation}
where the chemical potential $\mu_\psi$ is tuned to enforce the constraint $\langle n_{i;\psi} \rangle = 6$ per site. Using the operator mappings defined in Eqs.~\ref{eq:f_operator} and~\ref{eq:psi_operator}, we identify the parameters of Eq.\ref{eqn:THFM_mf} with those in Eq.\ref{eq:ancilla_mf}, yielding $\mu_f = -U/4 + 2U(1-\kappa) + \mu$ and $\mu_\psi = U/4 + 2U(1-\kappa) + \mu$ in the regime where $g_\gamma \approx 1$. Furthermore,  for $\gamma(\mathbf{k}) = 0$ or $\gamma(\mathbf{k}) \ll U$, we show that the Gutzwiller projected wavefunction derived from the renormalized mean field theory is exactly equivalent to the ancilla wavefunction at leading order of $\gamma/U$ (see the Supplementary)

\begin{figure}
    \centering
    \includegraphics[width=0.95\linewidth]{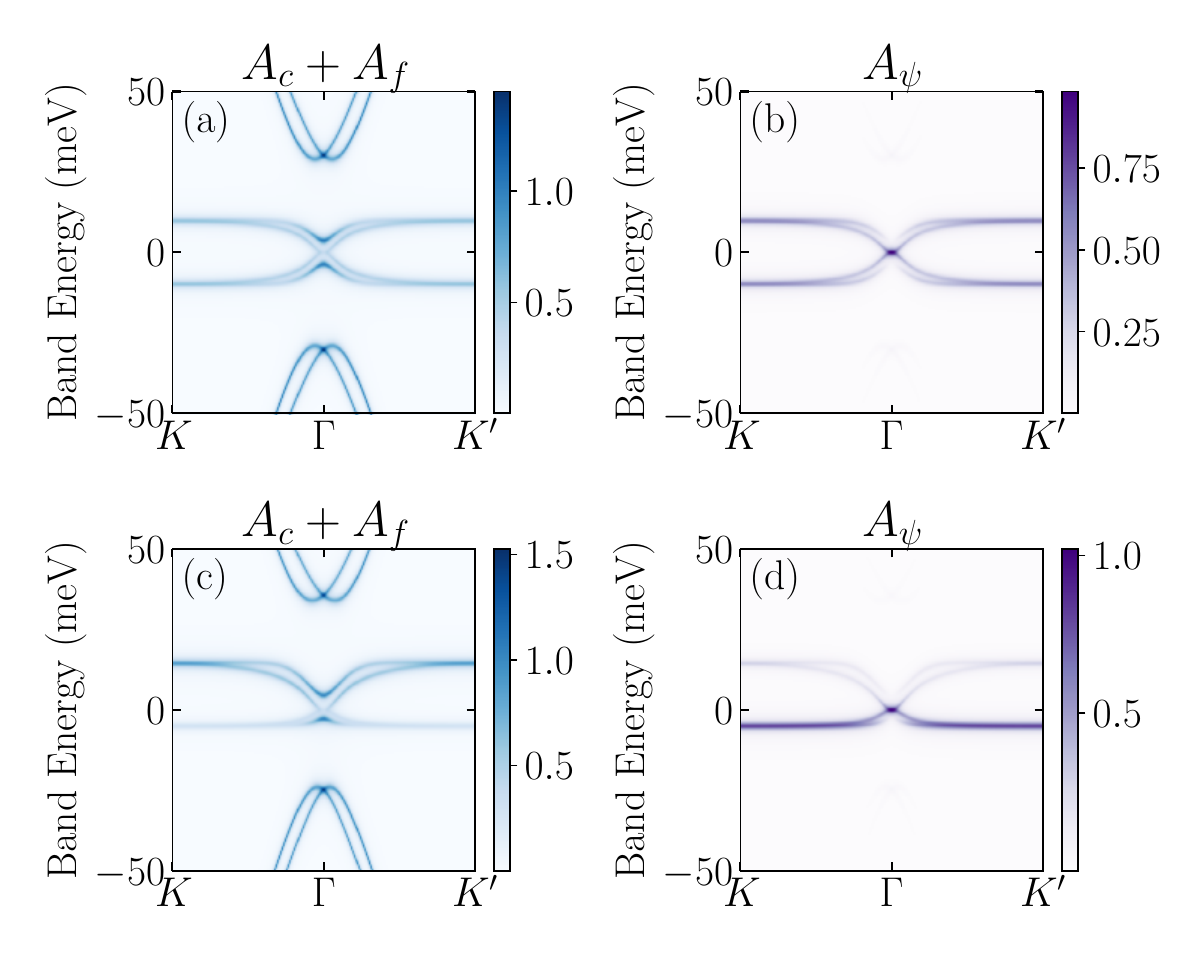}
    \caption{
    Band structure for semi-metals at charge neutrality point $\nu=0$ and integer filling $\nu=-2$.
    (a) Spectrum of physical electron $c+f$ and (b) auxiliary fermion $\psi$ at $U=20$meV, $\gamma=-38$meV and $\kappa=0.8$ for $\nu=0$. 
    (c) Spectrum of physical electron $c+f$ and (d) auxiliary fermion $\psi$ at $U=20$meV, $\gamma=-38$meV and a larger $\kappa = 1.1$ for $\nu=-2$. 
    }
    \label{fig:semimetal}
\end{figure}

\textbf{Mott semimetal} We have described an insulator at $\nu=-2$. We can perform similar treatment for $\nu=0$ by keeping $f^{3+}, f^{4+}, f^{5+}$ states.  As shown in Fig.~\ref{fig:semimetal}, now we have a semimetal with a quadratic band touching dominated by $\psi$. A semimetal also exists at $\nu=-2$ when $\kappa$ is larger than a critical value $\kappa_c\approx 1$.

\textbf{Discussion} In our above model at $\nu=-2$, we ignore  other doublon states  and only keep the spin-singlet doublon with angular momentum $L=0$. Strictly speaking, this is justified only when $J_A$ is very large. But we conjecture  that the physics is qualitatively similar at smaller $J_A$, as long as there is no symmetry breaking.  In the supplementary we consider a different model which includes two other spin-singlet doublon states with total angular momentum $L=\pm 2$. It turns out that the final spectrum does not change, consistent with the picture that the specific configuration of the spin state does not influence the charge sector.

\textbf{Conclusion} In summary, we propose a new theoretical method to analyze the THFM of TBG.  We obtain both ground state wavefunction and excitation spectrum of a mixed-valence Mott insulator at $\nu=-2$. The state features an exotic excitation in the hole doped side, which is orthogonal to the physical electron operator. Although we use the THFM, the physics discovered is quite distinct from the usual Kondo or heavy fermion systems. We conjecture that the composite excitation and momentum selective `Mott' gap may be universal in Anderson model with large hybridization, and not restricted to TBG. The  mean field theory developed here, complemented by the ancilla theory\cite{Zhao2025ancTBG} in our previous work, may be applied to analyze these models beyond TBG. In this work we keep $f^{1+}, f^{2+}, f^{3+}$ states at $\nu=-2$.  In our current framework, there is a cheap on-site Cooper pair operator $\ket{s_i} \bra{t_i} $, which can avoid the cost of Hubbard $U$.   In a subsequent work we plan to extend the model by including also the holon $f^{0+}$ state and discuss the superconductor upon hole doping, following the spirit of the `kinetic superconductor' recently discussed in bilayer nickelate\cite{yang2024strong,oh2024high}.

\textbf{Acknowledgement}  This work was supported by the National Science Foundation under Grant No. DMR-2237031.

\bibliography{refs}

\appendix

\section{THFM with anti-Hund's coupling}

We now present the full expression of the THFM with the anti-Hund spin coupling as introduced in Eq.~\eqref{eqn:THFM_0}.  
We begin with the dispersive $c_1$ and $c_2$ bands, described by
\begin{equation}\label{eq:THFM_c12}
\begin{aligned}
        H_0^{(c_1,c_2)} = &v_\star\sum_{\mathbf{k}}\left( c^\dagger_{1,\mathbf k}
        \tau_z\left( k_x\sigma_0+\mathrm{i}k_y\sigma_z\right) c_{2,\mathbf{k}}+\mathrm{h.c.}\right) \\
        &+\sum_{\mathbf{k}}c^\dagger_{2,\mathbf{k}}M\sigma_x c_{2,\mathbf{k}}, \\
\end{aligned}
\end{equation}
where the band width $v_\star |\mathbf{K}|$ is the largest energy scale in the model, and $M$ denotes the $\Gamma$-point splitting of the active flat bands in TBG.  
In all calculations presented, we take $v_\star = -4.544$ eV and $M = -3.7$ meV.

The hybridization between the dispersive $c_1$ band and the localized $f$ orbital is given by
\begin{equation}\label{eq:THFM_cf}
\begin{aligned}
        H^{(cf)}_0=&\frac{1}{\sqrt{N}}\sum_{\mathbf{k},i} e^{\mathrm{i}\mathbf{k}\cdot\mathbf{R}_i-\frac{k^2\lambda^2}{2}}\\
        &f^\dagger_{i}(\gamma\sigma_0 +v^\prime_\star\left(k_x\sigma_x+k_y\tau_z\sigma_y\right)\big)
        c_{1,\mathbf{k}}+\mathrm{h.c.},\\
\end{aligned}
\end{equation}
where the momentum-dependent hybridization takes the form $\gamma(\mathbf{k}) = e^{-k^2\lambda^2/2}\left(\gamma\sigma_0 +v^\prime_\star(k_x\sigma_x+k_y\tau_z\sigma_y)\right)$ 
and decays exponentially at large $|\mathbf{k}|$ due to the Wannier envelope.  
We set $v'_\star = -1.702$ eV and $\lambda = 0.3375$ (in units of the moiré lattice constant), 
while $\gamma=-38$ meV characterize the remote band gap at the $\Gamma$ point.
When we tune the hybridization strength, we multiply the hybridization matrix $\gamma(\mathbf k)$ by a coefficient $\gamma(\mathbf k)\rightarrow \lambda \gamma(\mathbf k)$, but we will always use the $\Gamma$ point gap to characterize the hybridization strength, as we did in Fig.~\ref{fig:kappa} (d). 

Finally, the anti-Hund spin interaction between the $f$ orbitals is modeled as
\begin{equation}\label{eqn:antiHundH}
\begin{split}
    H_{J_A}^{(f)}=&\frac{J_A}{4}\sum_{i,\mu,\rho}\left(f^\dagger_{i;K}\left(\sigma_\mu s_\rho\right)f_{i;K} \right)\left(f^\dagger_{i;K^\prime}\left(\sigma_\mu s_\rho\right)f_{i;K^\prime}\right)\\
    &+\frac{J^\prime_A}{4}\sum_{i,\rho}
    \left(f^\dagger_{i;K}\left(\sigma_0 s_\rho\right)f_{i;K}\right)\left(f^\dagger_{i;K^\prime}\left(\sigma_0 s_\rho\right)f_{i;K^\prime}\right)\\
    &-\frac{J^\prime_A}{4}\sum_{i,\rho}\left(f^\dagger_{i;K}\left(\sigma_z s_\rho\right)f_{i;K}\right)\left(f^\dagger_{i;K^\prime}\left(\sigma_z s_\rho\right)f_{i;K^\prime}\right). 
\end{split}    
\end{equation} 
where $\mu,\rho = x,y,z$ label Pauli matrices acting on orbital ($\sigma$) and spin ($s$) spaces, respectively.  
We use $f_{i;K}$ and $f_{i;K'}$ to represents $f$ 4-spinors in the $K$ and $K'$ valleys.  
Here we target the sector with $\tau_z=s^2=0$. The basis for this sector is:
\begin{equation}
    \begin{split}
        \ket{1_i}=&\frac{1}{2} \left(f^\dagger_{i;+K\uparrow} f^\dagger_{i;-K'\downarrow} 
    - f^\dagger_{i;+K\downarrow} f^\dagger_{i;-K'\uparrow} \right)\ket{0},\\
    \ket{2_i}=&\frac{1}{2} \left(f^\dagger_{i;-K\uparrow} f^\dagger_{i;+K'\downarrow} 
    - f^\dagger_{i;-K\downarrow} f^\dagger_{i;+K'\uparrow} \right)\ket{0},\\
    \ket{3_i}=&\frac{1}{2} \left(f^\dagger_{i;+K\uparrow} f^\dagger_{i;+K'\downarrow} 
    - f^\dagger_{i;+K\downarrow} f^\dagger_{i;+K'\uparrow} \right)\ket{0},\\
    \ket{4_i}=&\frac{1}{2} \left(f^\dagger_{i;-K\uparrow} f^\dagger_{i;-K'\downarrow} 
    - f^\dagger_{i;-K\downarrow} f^\dagger_{i;-K'\uparrow} \right)\ket{0}.\\
 \end{split}
\end{equation}
The Hamiltonian Eq.~\ref{eqn:antiHundH} projected onto this 2-particle basis can be written as:
\begin{equation}
\begin{pmatrix}
-J^\prime_A & -J_A & 0 & 0\\
-J_A & -J^\prime_A & 0 & 0 \\
0 & 0 & -J_A & 0 \\
0& 0& 0 & -J_A
\end{pmatrix}.
\end{equation}
For $J_A > 0$ and $J_A^\prime > 0$, the lowest energy state is the symmetric combination $\ket{d_i} = \tfrac{1}{\sqrt{2}}(\ket{1_i} + \ket{2_i})$, which has total angular momentum $L = 0$. In contrast, when $J_A > 0$ and $J_A^\prime < 0$, the lowest-energy manifold is twofold degenerate, consisting of the states $\ket{3_i}$ and $\ket{4_i}$ with angular momentum $L = \pm 2$, respectively. We denote these two states as $\ket{d_{1;i}}$ and $\ket{d_{2;i}}$. We discuss the case with $L = 0$ in the main text, and the $L = \pm 2$ case in Sec.~\ref{secsupp:dwave}.
\section{Details on the effective model and renormalized mean field theory}

\subsection{Relevant $f$ valences at $\nu=-2$}

For the $\nu=-2$ insulator, we are mostly interested in the $f^{1+}$, $f^{2+}$ and $f^{3+}$ valences at each AA site.
Specifically, when the anti-Hund's spin interaction is strong and $J_A>0, J^\prime_A>0$, 
the doubly occupied $f^{2+}$-orbital will be restricted to a single non-degenerate ground state of Eq.~\eqref{eqn:antiHundH} given by
\begin{equation}
\begin{aligned}
    |d_i\rangle =&\frac{1}{2} (f^\dagger_{i;+K\uparrow} f^\dagger_{i;-K'\downarrow} 
    - f^\dagger_{i;+K\downarrow} f^\dagger_{i;-K'\uparrow} \\
    &+ f^\dagger_{i;-K\uparrow} f^\dagger_{i;+K'\downarrow} 
    - f^\dagger_{i;-K\downarrow} f^\dagger_{i;+K'\uparrow}) |0\rangle~.
\end{aligned}
\end{equation}
where each flavor $f_{i;a\eta s}$ is paired with its time-reversal partner $f_{i;\bar a\bar\eta\bar s}$.

With a finite hybridization term $f^\dagger c_1$, 
the single occupied $f^{1+}$ and triply occupied $f^{3+}$ states will be excited.  
For the singly occupied state, we consider  
\begin{equation}
    |s_{i;a\eta s}\rangle = 2f_{i;\bar a\bar \eta\bar s}|d_i\rangle = -s f^\dagger_{i;a\eta s}|0\rangle~.  
\end{equation}
where $s = \pm 1$ for spin up and spin down.
It is straightforward to verify that
\begin{equation}
    \langle d_i|f^\dagger_{i;\bar a\bar\eta\bar s}|s_{i;a\eta s}\rangle =1/2~, 
\end{equation}

For the triply occupied sector, it suffices to consider the subspace generated by $f^\dagger_{i;a\eta s} |d_i\rangle$.  
Since 
\begin{equation}
    \langle d_i| f^{}_{i;a\eta s} f^\dagger_{i;a\eta s}|d_i\rangle = 3/4~, 
\end{equation}
we define the normalized triply occupied state as
\begin{equation}
    t^\dagger_{i;a\eta s} = \frac{2}{\sqrt{3}} f^\dagger_{i;a\eta s} |d_i\rangle~, 
\end{equation}
which yields
\begin{equation}
    \langle t_{i;a\eta s} | f^\dagger_{i;a\eta s} | d_i \rangle = \frac{\sqrt{3}}{2}.
\end{equation}

Note that the eight triple occupied states $|t_{i;a\eta s}\rangle$ defined above are not, in general, eigenstates of the spin interaction $H_{J_A}^{(f)}$ in Eq.~\eqref{eqn:antiHundH} and will be further split when $J_A$ is large.  
However, in this work we focus on the regime with moderate $J_A$ and neglect such splittings for simplicity.  
The primary role of the $J_A$ term is then to select the pairing channel responsible for the doubly occupied ground state.

We now project the localized $f$ orbital into subspace spanned by the relevant valences $|s\rangle, |d\rangle $ and $|t\rangle$:
\begin{equation}
    P_G = \otimes_i \left(|d_i\rangle\langle d_i| + \sum_{\alpha}|s_{i;\alpha}\rangle \langle s_{i;\alpha}|+\sum_\alpha|t_{i;\alpha}\rangle \langle t_{i;\alpha}|\right)~,
\end{equation}
After which
\begin{equation}
    P_G f^\dagger_{i;a\eta s}P_G = \frac{1}{2} |d_i\rangle \langle s_{i;\bar a\bar \eta\bar s}| +  \frac{\sqrt{3}}{2} |t_{i;a\eta s}\rangle\langle d_i|.
\end{equation}

Then we formulate a simple effective model by restricting to a subspace imposed by $P_G$. We map $\prod_i|d_i\rangle \rightarrow |0\rangle$, such that the product state of the $f^{2+}$ singlets $\otimes|d_i\rangle$ is treated as the vacuum.  Then we introduce two fermionic operators to create excitations on top of this vacuum:
\begin{equation}
    s^\dagger_{i;\alpha} = |s_{i;\alpha}\rangle \langle d_i|, \quad
    t^\dagger_{i;\alpha} = |t_{i;\alpha}\rangle \langle d_i|, 
\end{equation}
where the $s$ and $t$ operators should additionally satisfy the constraint 
\begin{equation}
    \sum_{a\eta s} s^\dagger_{i;a\eta s} s^{}_{i;a\eta s}+
    \sum_{a\eta s} t^\dagger_{i;a\eta s} t^{}_{i;a\eta s} \leq 1
\end{equation}
This constraint is very similar to that of the $t-J$ model where $\sum_\sigma c^\dagger_{i\sigma}c^{}_{i\sigma}\leq 1$. 
Finally, the original $f$ operator can be decomposed as two new kinds of fermions in the restricted Hilbert space as
\begin{equation}\label{eqn:fst}
    P_G f^\dagger_{i;a \eta s}P_G = \frac{1}{2} s_{i;\bar{a}\bar\eta\bar{s}} + \frac{\sqrt{3}}{2} t^\dagger_{i;a\eta s}. 
\end{equation}

We can also rewrite the Hamiltonian \eqref{eqn:THFM_0} into the restricted Hilbert space. 
While the $c$ electron part $P_GH^{(c_1,c_2)}_0P_G = H^{(c_1,c_2)}_0$ remain invariant, and the hybridization part $P_GH_0^{(cf)}P_G$ can be obtained by simply replacing the $f^\dagger_{i;a\eta s}$ operators as in Eq.~\eqref{eqn:fst},  
the interaction part should be treated more carefully since $P_Gf^\dagger_{i;a\eta s}f^{}_{i;a\eta s}P_G \neq P_Gf^\dagger_{i;a\eta s}P_Gf^{}_{i;a\eta s}P_G$. 

On the other hand, we should calculate exactly the interaction energy for each valences $f^{1+}, f^{2+}, f^{3+}$. 
The Hubbard part of these energies with a finite $\kappa$ at $\nu=-2$ are
$E_{1+}^f = (3-2\kappa)^2U/2$, $E_{2+}^f=(2-2\kappa)^2U/2$ and $E_{3+}^f=(1-2\kappa)^2U/2$. 
Note that $s^\dagger_{i;a\eta s}$ and $t^\dagger_{i;a\eta s}$ operators create $|s_{i;a\eta s}\rangle$ and $|t_{i;a\eta s}\rangle$ states from the vacuum $|d_i\rangle$, and only the relative energies compared to the $d$ states have physical meaning. 
We therefore write the Hubbard interaction in the projected Hilbert space as: 

\begin{widetext}

\begin{equation}
\begin{aligned}
    P_G (H_{i;\mathrm{int}}-\mu N_f)P_G = &
    P_G ((E_{1+}^f-\mu) \sum_{a\eta s}|s_{i;a\eta s}\rangle\langle s_{i;a\eta s}| + (E_{2+}^f-2\mu)|d_i\rangle \langle d_i| + (E_{3+}^f-3\mu) \sum_{a\eta s}|t_{i;a\eta s}\rangle\langle t_{i;a\eta s}|)P_G\\
    =& P_G\left(E_{2+}^f-2\mu + E_s \sum_{a\eta s} s^\dagger_{i;a\eta s} s^{}_{i;a\eta s} + E_t\sum_{a\eta s}t^\dagger _{i;a\eta s} t^{}_{i;a\eta s}\right)P_G~,
\end{aligned}
\end{equation}
where 
\begin{equation}
    E_s = (E_{1+}^f-\mu) - (E_{2+}^f-2\mu) = U/2 + 2U(1-\kappa) + \mu, 
\end{equation}
\begin{equation}
    E_t = (E_{3+}^f-3\mu) - (E_{2+}^f-2\mu) = U/2 - 2U(1-\kappa) - \mu. 
\end{equation}
In derivation, we used the relation 
\begin{equation}
    P_G|d_i\rangle \langle d_i| P_G= P_G \left(1-\sum_{a\eta s}|s_{i;a\eta s}\rangle \langle s_{i;a\eta s}|-\sum_{a\eta s}|t_{i;a\eta s}\rangle \langle t_{i;a\eta s}|\right)P_G. 
\end{equation}

\end{widetext}

\subsection{Renormalized Mean-Field theory}

To solve the restricted Hamiltonian $P_G H_0P_G$ with constraints $n_{i;s}+n_{i;t}\leq 1$, one way is to introduce a boson $d$ as slave particle for the vacuum, similar to the familiar slave boson theory of the  $t$-$J$ model. 
Instead, here we try to use a simpler renormalized mean-field theory (RMFT), which is an approximation method to effectively add the Gutwillzer projection effect without really doing variational Monte Carlo calculation. 

In the standard RMFT of $t$-$J$ model, one counts the probability of hopping $-t\sum_\sigma c^\dagger_{i\sigma}c^{}_{j\sigma}$ before and after applying the projection to $\sum_\sigma c^\dagger_{i\sigma}c^{}_{i\sigma}\leq 1$ \cite{zhang_rmft_1988}. 
In our present model, we need do similar treatment of the hybridization terms $P_G s_{i;\bar a\bar\eta\bar s} c^{}_{\mathbf k;a\eta s}P_G$ and $P_G t^\dagger_{i;a\eta s} c_{\mathbf k;a\eta s}P_G$. 
The relevant probability can be estimated by counting the fraction of randomly selected pairs of states from the Hilbert space that yield a non-zero matrix element.

Without any projection, the probabilities are
\begin{equation}
    p_{s,c}^0 = \sqrt{(1-n_{i;s;\bar a\bar \eta\bar s})n_{i;s;\bar a\bar\eta\bar s} (1-n_{\mathbf k;c;a\eta s})n_{\mathbf k;c;a\eta s}}~,
\end{equation}
and 
\begin{equation}
    p_{t,c}^0 = \sqrt{n_{i;t;a\eta s}(1-n_{i;s;a\eta s}) (1-n_{\mathbf k;c;a\eta s})n_{\mathbf k;c;a\eta s}}~.
\end{equation}

On the other hand, within the  projected Hilbert space, the $s^\dagger_{i;\alpha\eta s}$ and $t^\dagger_{i;\alpha\eta s}$ operator are non zero only when acting on the doubly occupied state $|d_i\rangle$, 
which gives the probability: 
\begin{equation}
    p_{s,c} = \sqrt{(1-n_{i;s}-n_{i;t})n_{i;s;\bar a\bar\eta \bar s} (1-n_{\mathbf k;c;a\eta s})n_{\mathbf k;c;a\eta s}}~,
\end{equation}
and 
\begin{equation}
    p_{t,c} = \sqrt{n_{i;t; a\eta s} (1-n_{i;s}-n_{i;t})(1-n_{\mathbf k;c;a\eta s})n_{\mathbf k;c;a\eta s}}~.
\end{equation}

The hopping term should be renormalized as the fraction between these two probabilities: 
\begin{equation}
    g_\gamma^s = \frac{p_{s,c}}{p^0_{s,c}} = \sqrt{\frac{1-n_{i;s}-n_{i;t}}{1-n_{i;s;\bar a\bar \eta \bar s}}} = \sqrt{\frac{n_d}{1-n_s/8}}, 
\end{equation}
and 
\begin{equation}
    g_\gamma^{t} = \frac{p_{t,c}}{p^0_{t,c}} = \sqrt{\frac{1-n_{i;s}-n_{i;t}}{1-n_{i;t; a \eta s}}} = \sqrt{\frac{n_d}{1-n_t/8}}~, 
\end{equation}
where $n_d = 1-n_s-n_t$ is the expectation value of $\sum_i|d_i\rangle\langle d_i|$
In realistic calculation, the denominator $1-n_s/8$ and $1-n_t/8$ is always found close to 1, so for simplicity we would take $g_\gamma^s = g_\gamma^t = g_\gamma = \sqrt{1-n_s-n_t}=\sqrt{n_d}$.  
The factor will also appear in the calculation of the spectrum function $A_f$ and $A_\psi$, where the mean-field spectrum should be multiplied by a factor $g_\gamma^2 = n_d$ to estimate the physical spectrum weight. 

\section{Charge neutrality point $\nu=0$.}

Although our focus is mainly on $\nu=-2$ correlated insulator in this work, our method can be easily generalized to other even integer fillings $\nu=0,\nu=+2$. 
Here we show the results of the charge neutrality point $\nu=0$. 
We now consider three different valences of the $f^{3+}$, $f^{4+}$ and $f^{5+}$ orbital, which are denoted as $|t_\alpha\rangle$, $|q\rangle$, and $|p_\alpha\rangle$. 
Similar to our treatment of $\nu=-2$, we only include the spin-singlet state (the ground state of $H_{J_A}$ for the $f^{4+}$ valence) and denote it as $\ket{q_i}$. By treating the $\otimes |q_i\rangle$ state as the vacuum, we can do a similar projection:
\begin{equation}
    P_G^{\nu=0} f^\dagger_{i;a\eta s}P^{\nu=0}_G = \frac{1}{\sqrt{2}} t_{i;\bar a\bar\eta \bar s} + \frac{1}{\sqrt{2}} p^\dagger_{i;a\eta s},
\end{equation}
And we can introduce an auxiliary fermion $\psi$ as : 
\begin{equation}
    \psi^\dagger_{i;a\eta s} = A\{f^\dagger_{i;a\eta s}, n_{i;f}\} - Bf^\dagger_{i;a\eta s}~,
\end{equation}
with $A=1, B=8$ now. 
Under the same projections, we get: 
\begin{equation}
    P^{\nu=0}_G\psi_{i;a\eta s}^\dagger P^{\nu=0}_G = -\frac{1}{\sqrt{2}} t_{i;\bar a\bar \eta \bar s}^{} + \frac{1}{\sqrt{2} }p^\dagger_{i;a\eta s}. 
\end{equation}
The corresponding spectrum for $U=20$ meV and $\gamma=-38$ meV are shown in Fig.~\ref{fig:semimetal} (a) and (b) in the main text, 
which is consistent with earlier results \cite{Zhao2025ancTBG} by the ancilla fermion method. 

For odd integer fillings $\nu=\pm1,\pm3$, the current slave particle method is not applicable due to the absence of non-degenerate paired state at these fillings. 
However, we believe similar ansatz also exist when the temperature is high enough to kill the local moment information, or when the local moments form spin liquid states without long range spin order. 
Directly applying the ancilla fermion method in Ref.~\cite{Zhao2025ancTBG} is more convenient in those cases.  

\section{Another model with d-wave doublons at $\nu=-2$}\label{secsupp:dwave}

Here we also try to discuss the case where the doubly occupied state is favored to be in other channels, e.g. the d-wave channel. 
In this case, there is a double degeneracy of the d-wave ground state for the $f^{2+}$ valence:
\begin{equation}
\begin{aligned}
    |d_{1;i}\rangle = & \frac{1}{\sqrt{2}}(f_{i;+K\uparrow}^\dagger f_{i;+K'\downarrow}^\dagger-
    f_{i;+K\downarrow}^\dagger f_{i;+K'\uparrow}^\dagger)|0\rangle\\
    |d_{2;i}\rangle = & \frac{1}{\sqrt{2}}(f_{i;-K\uparrow}^\dagger f_{i;-K'\downarrow}^\dagger-
    f_{i;-K\downarrow}^\dagger f_{i;-K'\uparrow}^\dagger)|0\rangle~.
\end{aligned}
\end{equation}

Here $d_1$ and $d_2$ have anular momentum $L=+2$ and $L=-2$ respectively. For simplicity we will use the slave boson conventions and view $\ket{d_{a;i}}=d^\dagger_{a;i}\ket{0}$ for $a=1,2$. $d_1,d_2$ are two different slave bosons. Then we have the projected electron operator:
\begin{equation}
\begin{aligned}
    P_G^d f^\dagger_{i;+ \eta s}P_G^d =& \frac{s}{\sqrt{2}}d_{1;i}^\dagger s_{i;+\bar\eta\bar{s}} + \frac{1}{\sqrt{2}} t^\dagger_{1;i;+\eta s} d_{1;i} + t^\dagger_{2;i;+\eta s} d_{2;i}\\
    P_G^d f^\dagger_{i; -\eta s}P_G^d = & \frac{s}{\sqrt{2}}d_{2;i}^\dagger s_{i;-\bar\eta\bar{s}} + \frac{1}{\sqrt{2}} t^\dagger_{2;i;-\eta s} d_{2;i} + t^\dagger_{1;i;-\eta s} d_{1;i}\\
\end{aligned}
\end{equation}
where $t^\dagger_1, t^\dagger_2$ create two different kinds of $f^{3+}$ states orthogonal to each other, $s^{\dagger}$ creates the $f^{1+}$ states. 

We can assume that $d_1$ and $d_2$ are always condensed and treat $\langle d_1 \rangle, \langle d_2 \rangle$ as variational parameters. The above parton construction actually has two independent U(1) gauge symmetry associated with 
$d_{1;i} \rightarrow d_{1;i}e^{i \theta_{1;i}}$, 
$s_{i;+\eta s} \rightarrow s_{i;+\eta s}e^{i \theta_{1;i}}$, 
$t_{1;i;a\eta s} \rightarrow t_{1;i;a\eta s}e^{i \theta_{1;i}}$, 
and $d_{2;i} \rightarrow d_{2;i}e^{i \theta_{2;i}}$, 
$s_{i;-\eta s} \rightarrow s_{i;-\eta s}e^{i \theta_{2;i}}$, 
$t_{2;i;a\eta s} \rightarrow t_{2;i;a\eta s}e^{i \theta_{2;i}}$, 
separately. 
$d_1$ and $d_2$ are related by the $C_2T$ symmetry.  We consider a symmetric ansatz with $\langle d_1 \rangle=\langle d_2 \rangle$.
In this case, we can combine the $t_1$ and $t_2$ particle and get:
\begin{equation}
    P_G^d f^\dagger_{i;a \eta s}P_G^d = \frac{s}{2} \langle d \rangle^* s_{i;a \bar\eta\bar{s}} + \frac{\sqrt{3}}{2} \langle d \rangle t^\dagger_{i;a\eta s} 
\end{equation}
where we consider the ansatz $  \langle d_{i;1}\rangle = \langle d_{i;2}\rangle=\frac{1}{\sqrt{2}} \langle d \rangle$. We expect $\langle d \rangle=\sqrt{1-n_s-n_t}$.
We also introduce a new $f^{3+}$ state created by $t^\dagger_{i;a\eta s}=\sqrt{\frac{1}{3}}t^\dagger_{1;i; a \eta s}+\sqrt{\frac{2}{3}}t^\dagger_{2;i; a \eta s}$.

This is the same as the result for the model in the main text, which includes only the $s$-wave doubon.  Thus we expect exactly the same single particle spectrum. We also emphasize that this state is symmetric under both the $C_2T$ symmetry and the $C_3$ rotation symmetry. Naively $\langle d_1 \rangle =\langle d_2 \rangle \neq 0$ breaks the $C_3$ rotation symmetry, but the final phase after the projection is actually symmetric by combining the mean field $C_3$ rotation with a gauge transformation, in the spirit of projective symmetry.

We expect the correlated insulator  to remain the same as long as the condensation is ``uniform'' within all the flavors. 
We therefore conjecture that our correlated insulator result is quite general as long as no spin-order is seen (such as valley polarization) in the two-particle ground state. 

\begin{figure}[t]
    \centering
    \includegraphics[width=0.98\linewidth]{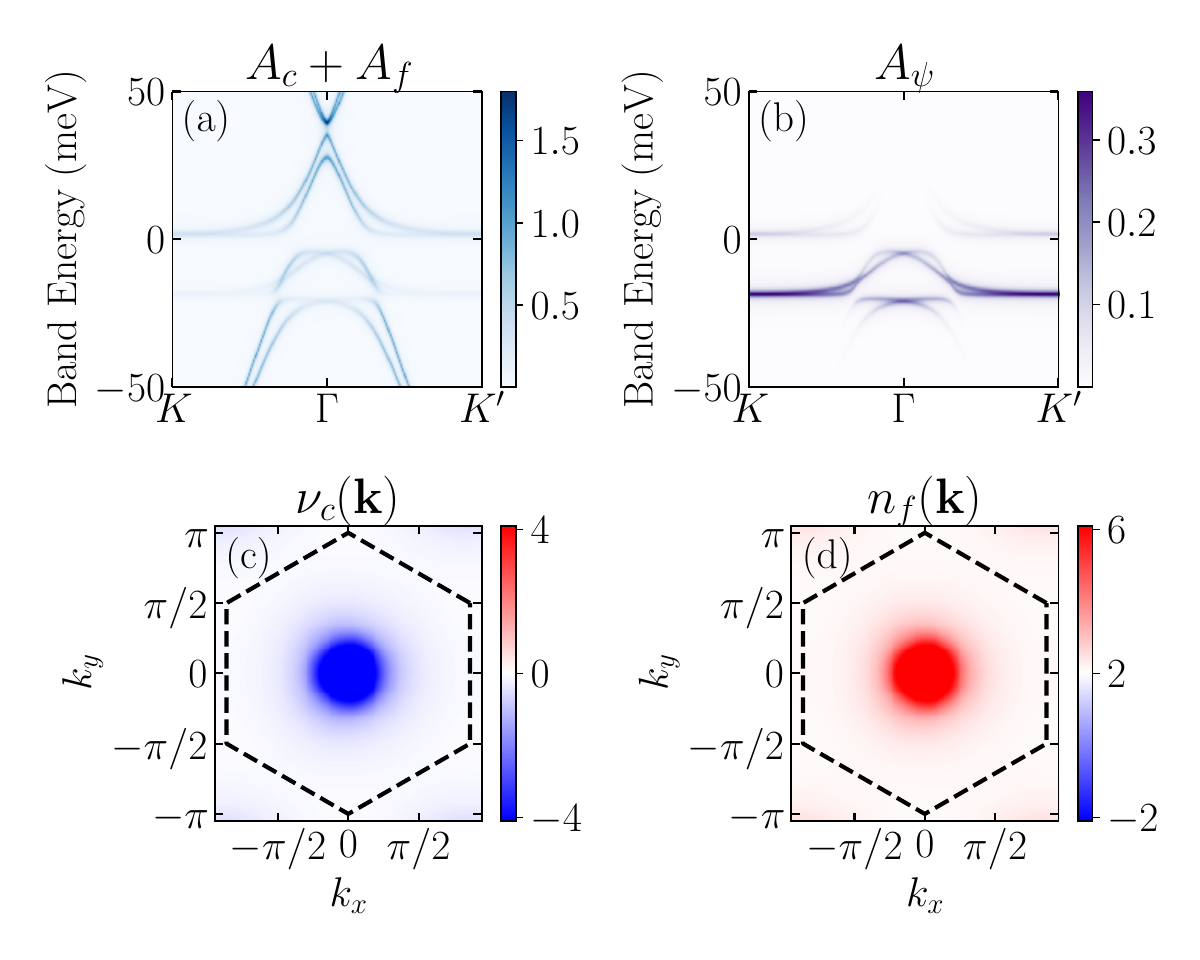}
    \caption{Typical mean field band structure calculated at $U=20$meV, $\gamma=-38$meV, $\kappa=0.0$. (a) The total physical electron spectrum. 
    (b) Spectrum of the auxiliary fermion $\psi$. 
    (c) and (d) shows the momentum space density distribution $\nu_c(\mathbf{k})$ of $c$ orbital and $n_f(\mathbf{k})$ of $f$ orbital.  
    }
    \label{fig:spectrum_kappa0}
\end{figure}

\begin{figure}[t]
    \centering
    \includegraphics[width=0.98\linewidth]{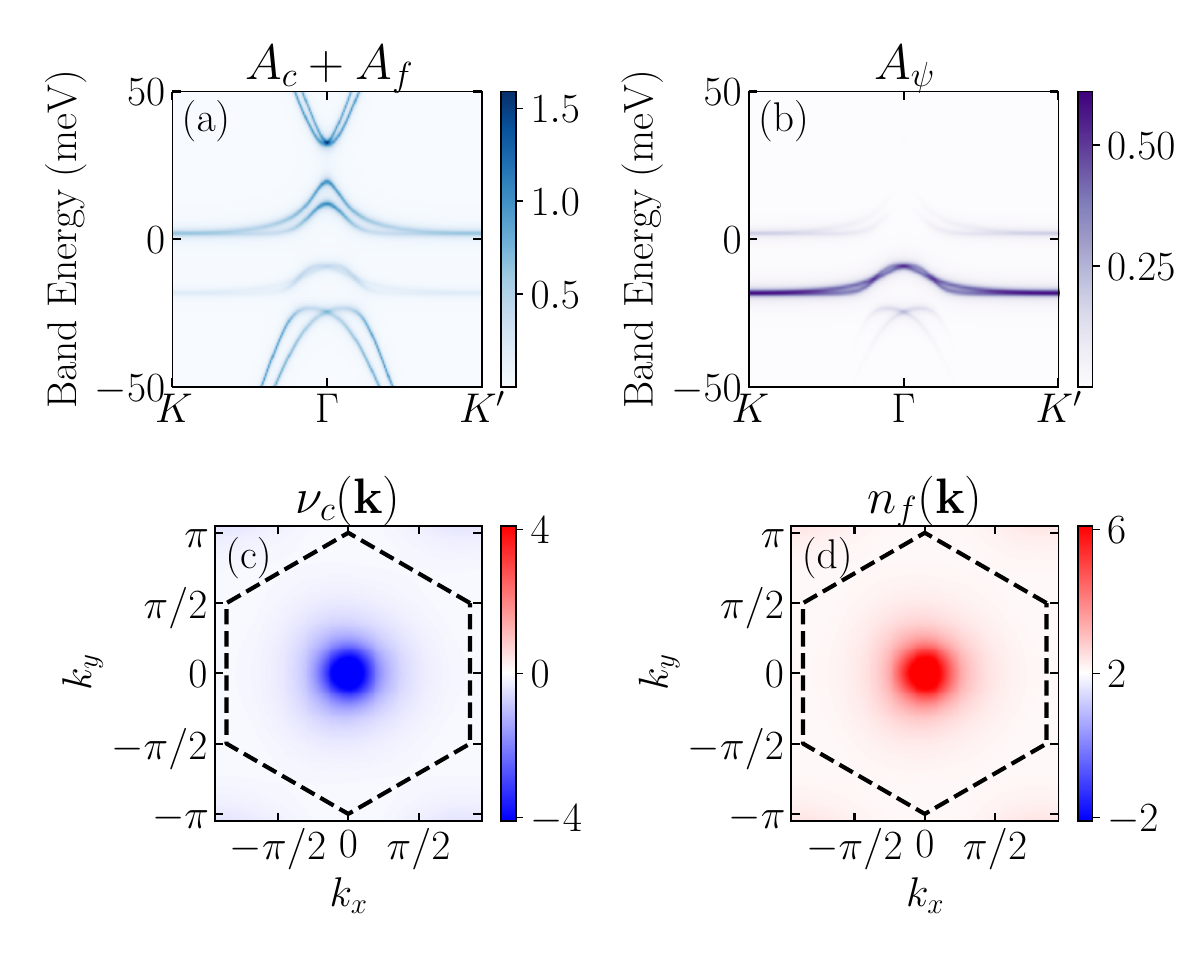}
    \caption{Typical mean field band structure calculated at $U=20$meV, $\gamma=-38$meV, $\kappa=0.4$. (a) The total physical electron spectrum. 
    (b) Spectrum of the auxiliary fermion $\psi$. 
    (c) and (d) shows the momentum space density distribution $\nu_c(\mathbf{k})$ of $c$ orbital and $n_f(\mathbf{k})$ of $f$ orbital. 
    }
    \label{fig:spectrum_kappa04}
\end{figure}

\section{More results for the slave-particle mean field calculation. }

\subsection{Spectrum for smaller $\kappa$}

In Figs.~\ref{fig:spectrum_kappa0} and \ref{fig:spectrum_kappa04}, we further show the mean-field spectrum calculated at smaller effective attractive potential $\kappa = 0.0$ and $\kappa = 0.4$. 
As a result of smaller $\kappa$,
there are more holes doped into the $c$ band than into the $f$ band. 
The average valence $f$ orbitals for $\kappa=0.0$ and $\kappa = 0.4$ are $f^{2.7+}$ and $f^{2.5+}$, respectively. 

Due to a different chemical potential, the upper Hubbard band now has a different band shape compared to that of $\kappa=0.8$. 
On the other hand, the lower Hubbard band is less sensitive to $\kappa$, and is always dominated by the auxilliary $\psi$ fermion.

\begin{figure}[t]
    \centering
    \includegraphics[width=0.95\linewidth]{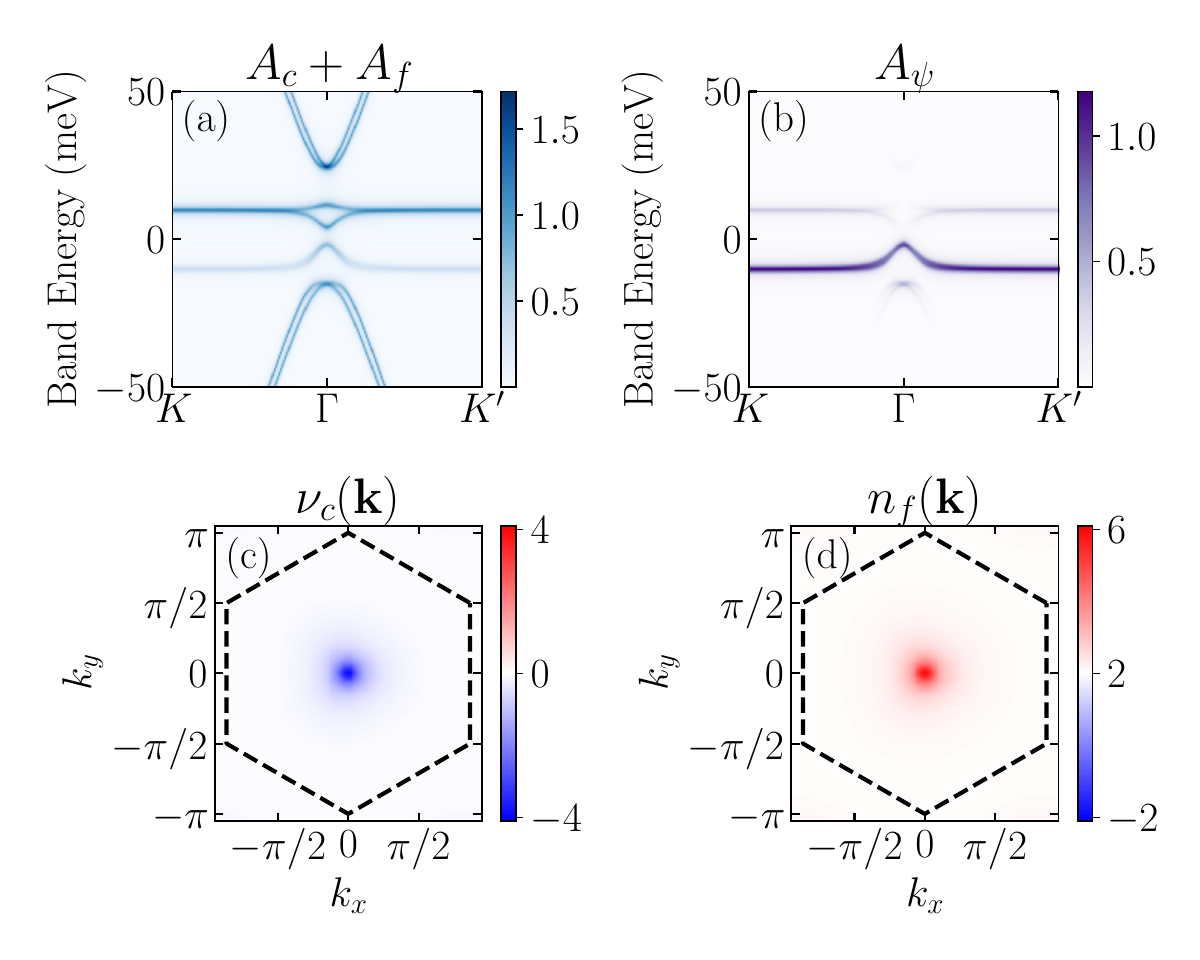}
    \caption{Mean field band structure calculated at $U=20$meV, $\kappa=0.8$, but half the hybridization strength $\gamma=-19$ meV, $v'_\star = -0.851$ eV.
    (a) The total physical electron spectrum. 
    (b) Spectrum of the auxiliary fermion $\psi$. 
    (c) and (d) shows the momentum space density distribution $\nu_f(\mathbf{k})$ of $c$ orbital and $n_f(\mathbf{k})$ of $f$ orbital. 
    }
    \label{fig:spectrum_20}
\end{figure}

\begin{figure}[t]
    \centering
    \includegraphics[width=0.95\linewidth]{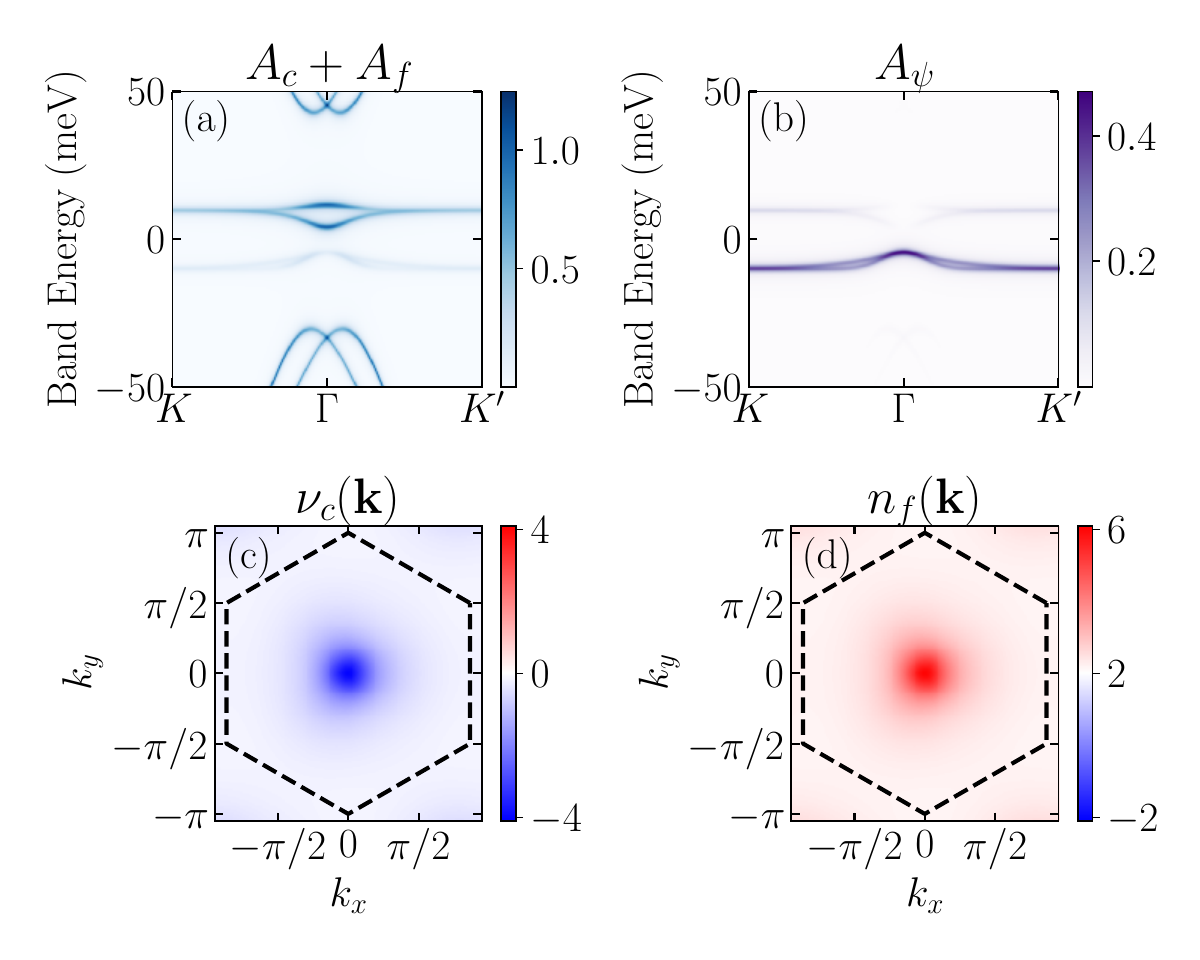}
    \caption{Mean field band structure calculated at $U=20$meV, $\kappa=0.8$, but double the hybridization strength $\gamma = -76$ meV,$v'_\star=-3.404$eV. 
    (a) The total physical electron spectrum. 
    (b) Spectrum of the auxiliary fermion $\psi$. 
    (c) and (d) shows the momentum space density distribution $\nu_f(\mathbf{k})$ of $c$ orbital and $n_f(\mathbf{k})$ of $f$ orbital.} 
    \label{fig:spectrum_80}
\end{figure}

\subsection{Spectra for different hybridization strength. }

In Figs.~\ref{fig:spectrum_20} and \ref{fig:spectrum_80}, we show the spectra at different hybridization strengths $\gamma$ for the $\nu=-2$ insulator. 
The overall band structures remain similar to that in Fig.~\ref{fig:spectrum} of the main text. 
The primary difference appears near the $\Gamma$ point: 
as $\gamma$ increases, a larger region of the Brillion zone deviates from the trivial flat Mott band. 

\begin{figure}[t]
    \centering
    \includegraphics[width=0.95\linewidth]{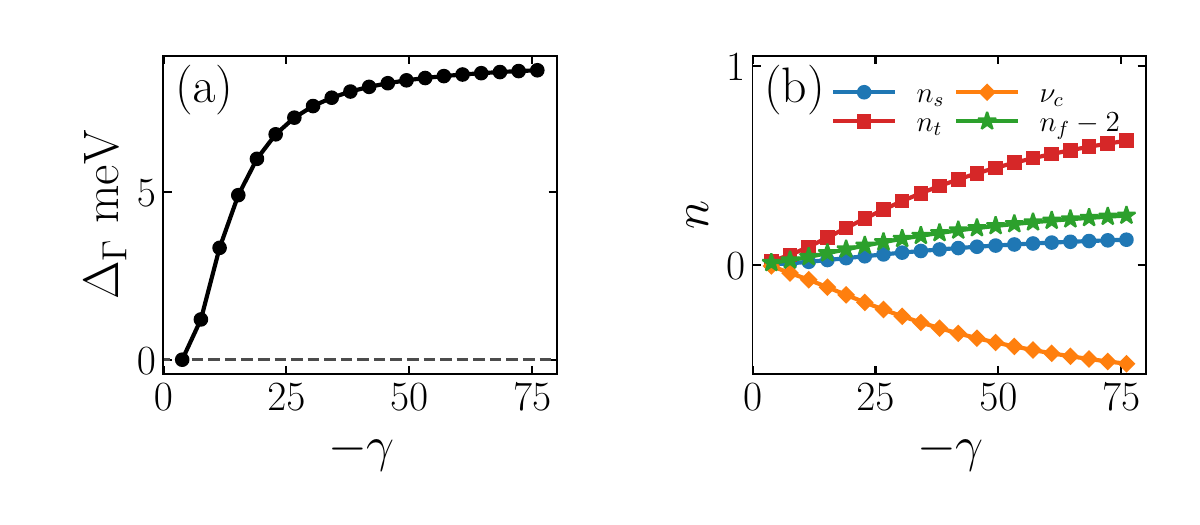}
    \caption{Band structure as a function of $\gamma$ with fixed interaction $U=20$ meV, $\kappa=0.8$. (a) $\Gamma$ point gap $\Delta_\Gamma$ as a function of $-\gamma$. 
    (b) shows the particle density $n_s$, $n_t$, $\nu_c$ and $n_f-2$ as a function of $-\gamma$. }
    \label{fig:gammas}
\end{figure}

In Fig.~\ref{fig:gammas} we further show the dependence of the $\Gamma$ point gap, and the particle densities as a function of $\gamma$. 
Here, $\gamma$ and $v'_\star$ are scaled by the same factor, but we use $\gamma$ as the $x$-axis variable.
At zero hybridization, no trion $t$ or singlon $s$ is excited, and the charge transfer gap $\Delta_\Gamma$ closes. 
On the other hand, with increasing $|\gamma|$, more and more $t$ and $s$ are excited and the $f$ orbital becomes more self-doped. 
As a result, the $\Gamma$ point gap increases from 0 and saturates to a constant.

\subsection{Doping the correlated insulator $\nu=-2$}

\begin{figure}[ht]
    \centering
    \includegraphics[width=0.95\linewidth]{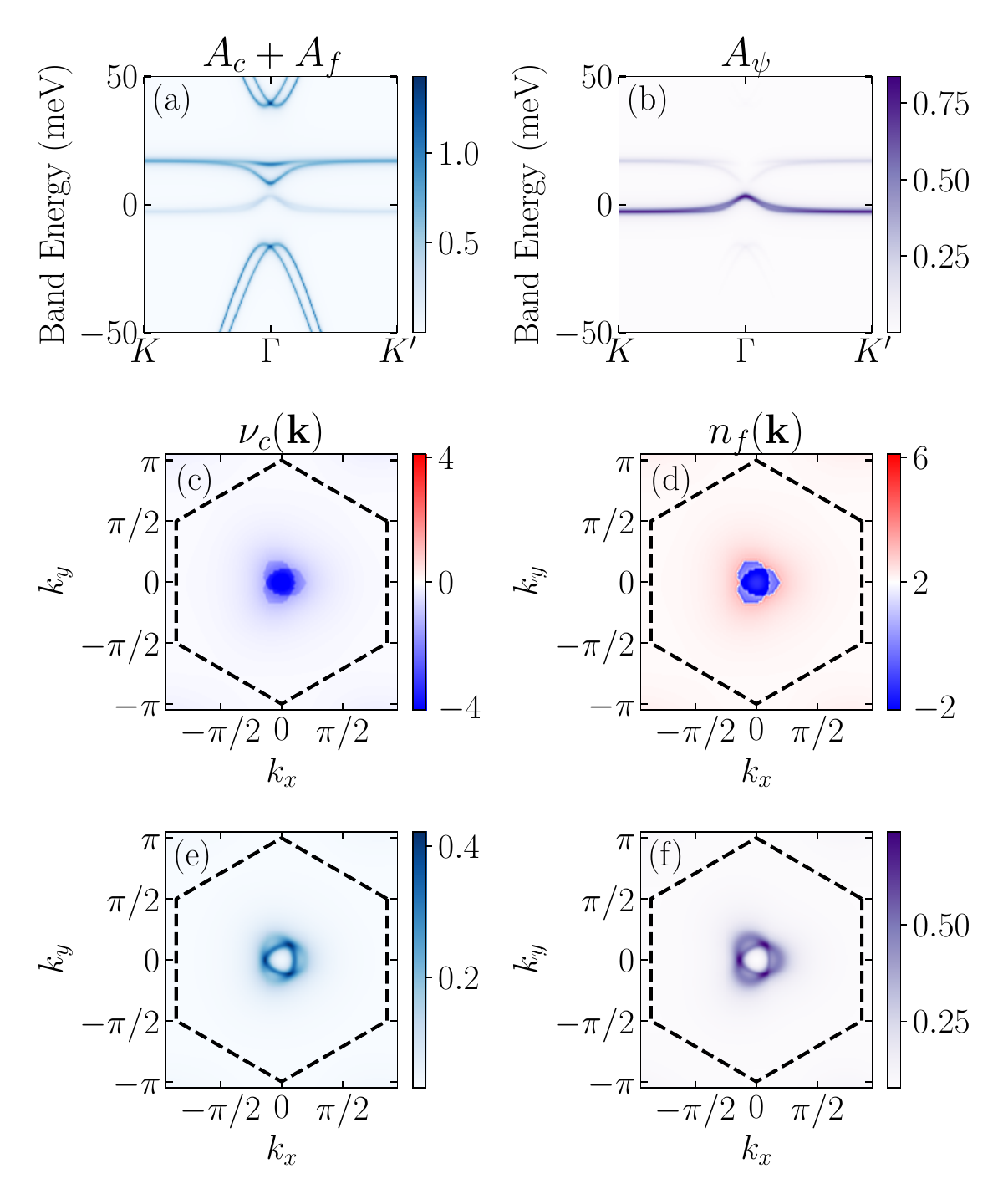}
    \caption{Mean field band structure calculated at $U=20$meV, $\gamma=-38$ meV, $\kappa=0.8$, and hole dopping level $\nu=-2-0.2$. 
    (a) The total physical electron spectrum. 
    (b) Spectrum of the auxiliary fermion $\psi$. 
    (c) and (d) shows the momentum space density distribution $\nu_f(\mathbf{k})$ of $c$ orbital and $n_f(\mathbf{k})$ of $f$ orbital.
    (e) and (f) shows the zero energy spectrum contributed from the physical observed component $A_c+A_f$ and from the ancilla fermion $A_\psi$, respectively. 
    } 
    \label{fig:spectrum_-2-x}
\end{figure}

We now consider lightly doping the correlated insulator at $\nu = -2$ into a metallic state. 
In Fig.~\ref{fig:spectrum_-2-x}, we present the spectral function for parameters $U = 20$~meV, $\gamma = -38$~meV, and doping level $\nu =-2- 0.2$. 
Consistent with the analysis of the insulating state, the low-energy spectral weight near the Fermi level is dominated by the auxiliary $\psi$ fermi.

\section{Equivalence between ancilla theory and renormalized mean field theory at small $\gamma(\mathbf{k})$}
In this section, we compare the ancilla wavefunction and Gutziwiller projected wavefunction, then prove that they are equivalent  while $\gamma(\mathbf{k})\ll U$. In the ancilla theory, we introduce two ancilla fermions $\psi$ and $\psi'$, considering the following ansatz:
\begin{equation}
    \ket{\Psi_\mathrm{ancilla}}=P_S\left(\ket{\mathrm{Slater}[c,f,\psi]}\otimes\ket{\Psi_{\psi'}}\right),
\end{equation}
where $P_S$ is a projection operator enforcing: (I) $n_{i;\psi}=4-\nu$; (II)$n_{i;\psi'}=4+\nu$; (III) the two local ancilla qubits $\psi_i$ and $\psi'_i$ form as SU(8) singlet at each site $i$. Here we target the filling $\nu=-2$ and choose the wavefunction of $\psi'$ to be:
\begin{equation}
  \begin{split}  \ket{\Psi_{\psi'}}=&\prod_{i=1}^{N_\mathrm{site}} \frac{1}{2}\big(\psi^{\prime\dagger}_{i;+K\uparrow}\psi^{\prime\dagger}_{i;-K'\downarrow}-\psi^{\prime\dagger}_{i;+K\downarrow}\psi^{\prime\dagger}_{i;-K'\uparrow}\\
  &+\psi^{\prime\dagger}_{i;-K\uparrow}\psi^{\prime\dagger}_{i;+K'\downarrow}-\psi^{\prime\dagger}_{i;-K\downarrow}\psi^{\prime\dagger}_{i;+K'\uparrow}\big)\ket{0}
  \end{split}
\end{equation}
due to large $J_A,J^\prime_A$. The charge sector is determined by the physical fermions $c_1,c_2,f$ and the first ancilla fermion $\psi$. This is taken as a Slater determinant $\ket{\mathrm{Slater}[c,f,\psi]}$ and dertermined by the following mean field Hamiltonian:
\begin{equation}\label{eqsupp:ancilla_mf} \begin{split}
    H_\mathrm{ancilla}=&H_0^{(c_1,c_2)}+H_0^{(c_1,f)}-\mu\left(N_{c_1}+N_{c_2}\right)\\
    &+\Phi\sum_\mathbf{k}\left(f^\dagger_{\mathbf{k}}\psi_{\mathbf{k}}+\mathrm{H.c.}\right)-\mu_f N_f-\mu_\psi N_\psi,
    \end{split}
\end{equation}
where $\mu_f$ is chosen to be $\mu$ in the previous study~\cite{Zhao2025ancTBG}. But in principle we can take other value of $\mu_f$. $\mu$ and $\mu_\psi$ are tuned to make $\braket{N_{c_1}}+\braket{N_{c_2}}+\braket{N_f}=2N_\mathrm{site}$ and $\braket{N_{\psi}}=6N_\mathrm{site}$.

In the renormalied mean field theory, we have two slave fermions $s$ and $t$, the ansatz is:
\begin{equation}
    \ket{\Psi_\mathrm{GP}}=P_G \ket{\mathrm{Gauss}[c,s,t]},
\end{equation}
where $P_G$ is the Gutziwiller projection operator enforcing $n_{i;s}+n_{i;t}\le 1$ at each site $i$. The Gaussian wavefuncion $\ket{\mathrm{Gauss}[c,s,t]}$ is determined by the following mean-field Hamiltonian:
\begin{equation}
    \begin{split}
    H_\mathrm{MF}=&H_0^{(c_1,c_2)}-\mu (N_{c_1}+N_{c_2}) \\&+\sum_i\left(E_s n_{i;s}+E_t n_{i;t}\right)+\sum_{\mathbf{k},i,\alpha,\beta}\Big(\frac{e^{\mathrm{i}\mathbf{k}\cdot\mathbf{R}_i}g_\gamma}{\sqrt{N_\mathrm{site}}}\\
    &\frac{s_{i;\bar{\alpha}}+\sqrt{3}t^\dagger_{i;\alpha}}{2}\gamma_{\alpha\beta}(\mathbf{k})c_{1;\mathbf{k};\beta}+\mathrm{H.c.}\Big),
    \end{split}
\end{equation}
where $g_\gamma=\sqrt{1-\braket{n_s}-\braket{n_t}}$, $\mu$ is tuned to make $\braket{N_{c_1}}+\braket{N_{c_2}}+\braket{N_t}-\braket{N_s}=0$. $E_s=U/2+2U(1-\kappa)+\mu$ and $E_t=U/2-2U(1-\kappa)-\mu$ as defined in the main text. To relate $H_\mathrm{ancilla}$ and $H_\mathrm{MF}$, here we define $\tilde{f}$ and $\tilde{\psi}$ as:
\begin{equation}
    \begin{split}
        \tilde{f}^\dagger_{i;\alpha}=&\frac{1}{2}s_{i;\bar{\alpha}}+\frac{\sqrt{3}}{2}t^\dagger_{i;\alpha},\\
        \tilde{\psi}^\dagger_{i;\alpha}=&-\frac{\sqrt{3}}{2}s_{i;\bar{\alpha}}+\frac{1}{2}t^\dagger_{i;\alpha}.
    \end{split}
\end{equation}
By replacing $s,t$ with $\tilde{f},\tilde{\psi}$, the mean-field Hamiltonian can be rewritten as:
\begin{equation}\label{eqsupp:HSPnewbasis}
    \begin{split}
    H_\mathrm{MF}=&H_0^{(c_1,c_2)}+g_\gamma H^{(c_1,\tilde{f})}-\mu(N_{c_1}+N_{c_2})\\
    &+\frac{\sqrt{3}U}{4}\sum_\mathbf{k}\left(\tilde{f}^\dagger_{\mathbf{k}}\tilde{\psi}_{\mathbf{k}}+\mathrm{H.c.}\right)-\frac{U}{2}N_{\tilde{\psi}}\\
    &+\left(\frac{U}{4}-2U(1-\kappa)-\mu\right)N_{\tilde{f}}\\
    &+\left(-\frac{U}{4}-2U(1-\kappa)-\mu\right)N_{\tilde{\psi}}.
\end{split}
\end{equation}
In the above mean field $H_\mathrm{MF}$, we can define $\mu_{\tilde{f}}=-U/4+2U(1-\kappa)+\mu$ and $\mu_{\tilde{\psi}}=U/4+2U(1-\kappa)+\mu$. With these definitions, the above mean field $H_{\mathrm{MF}}$ becomes very similar to $H_\mathrm{ancilla}$, up to two distinctions: (1) the term $H_0^{(c_1,\tilde{f})}$ is renormalized by a factor $g_\gamma$, and (2) $N_{\tilde{\psi}}$ is not fixed to $6N_{\mathrm{site}}$. Instead, $\mu_{\tilde{\psi}}$ is fixed to be $U/4+2U(1-\kappa)+\mu$.

In the special case that $\gamma(\mathbf{k})=0$, we have $\braket{n_s}=\braket{n_t}=0, g_\gamma=1$ and $\mu=0$. $c$ fermions decouple from $\tilde{f},\tilde{\psi}$ fermions. The part of $\tilde{f}$ and $\tilde{\psi}$ in the mean field Hamiltonian can be written as:
\begin{equation}
\begin{split}
    H^{(\tilde{f}\tilde{\psi})}_{\mathrm{MF}}=&\frac{\sqrt{3}U}{4}\sum_{\mathbf{k}}\left(\tilde{f}^\dagger_{\mathbf{k}}\tilde{\psi}_{\mathbf{k}}+\mathrm{H.c.}\right)-\frac{U}{4}\left(N_{\tilde{\psi}}-N_{\tilde{f}}\right)\\
    &+2U(1-\kappa)\left(N_{\tilde{\psi}}+N_{\tilde{f}}\right).
\end{split}
\end{equation}
In the parameter regime $3/4<\kappa<5/4$, we can solve that $n_{\tilde{\psi}}=6$, thereby recovering the constraint in the original ancilla theory. As a result, $H_{\mathrm{MF}}$ becomes exactly equivalent to $H_{\mathrm{ancilla}}$ under the identification $\Phi = \sqrt{3}U/4$ and $\mu_f = -\tfrac{U}{4} + 2U(1 - \kappa) + \mu$. We further show that the wavefunction $\ket{\Psi_{\mathrm{ancilla}}}$ and $\ket{\Psi_{\mathrm{GP}}}$ are identical. Moreover, this equivalence can be extended perturbatively to the case that $\gamma(\mathbf{k})\ne0$ but $\gamma(\mathbf{k})\ll U$.

\subsection{$\gamma(\mathbf{k})=0$}
We first consider the case with $\gamma(\mathbf{k})=0$. In the ancilla approach, $f$ strongly hybridizes with the first ancilla $\psi$ on each site $i$:
\begin{equation}\label{eqsupp:ancillawavefunction0}
\begin{split}
    \ket{\mathrm{Slater}[c,f,\psi]}_0=&\ket{\mathrm{Slater}[c_1,c_2]}_0\otimes\ket{\mathrm{Slater}[f,\psi]}_0,\\
    \ket{\mathrm{Slater}[f,\psi]}_0=&\prod_{i=1}^{N_\mathrm{site}}\prod_\alpha\left(\frac{1}{2}f^\dagger_{i;\alpha}-\frac{\sqrt{3}}{2}\psi^\dagger_{i;\alpha}\right)\ket{0}.
\end{split}
\end{equation}
Here the form of $\ket{\mathrm{Slater}[f,\psi]}$ is provided in the Ref.~\cite{Zhao2025ancTBG}. $\ket{\mathrm{Slater}[c_1,c_2]}_0$ is the Slater determinant of $c_1,c_2$ at charge neutrality.
After tensor producting the wavefunction of the second ancilla layer $\ket{\Psi_{\psi'}}$, the projection operator $P_S$ teleportates the state from $\psi'$ to $f$. The final ancilla wavefunction can be written as:
\begin{equation}
\begin{split}
    \ket{\Psi_\mathrm{ancilla}}_0=&\ket{\mathrm{Slater}[c_1,c_2]}_0\otimes \ket{\Psi_f},\\
    \ket{\Psi_f}=&\left(\ket{\Psi_{\psi'}}\ \mathrm{replacing}\ \psi' \mathrm{with} \ f\right).
\end{split}
\end{equation}
In the renormalized mean field theory, the Gutziwiller projected wavefunction is:
\begin{equation}
    \begin{split}
    \ket{\Psi_\mathrm{GP}}_0=&P_G\ket{\mathrm{Gauss}[c,s,t]}_0\\
    =&P_G\ket{\mathrm{Slater}[c,\tilde{f},\tilde{\psi}]}_0\\
    =&\ket{\mathrm{Slater}[c_1,c_2]}_0\otimes \prod_{i=1}^{N_\mathrm{site}}\ket{d_i},
    \end{split}
\end{equation}
where $\ket{d_i}$ is defined in Appendix A. Therefore $\ket{\Psi_\mathrm{GP}}=\ket{\Psi_\mathrm{ancilla}}$.

In addition to the equivalence of wavefunctions, we can show the projection operators $P_S$ and $P_G$ have the same effect. As shown in Ref.~\cite{Zhao2025ancTBG}, $P_S$ in the ancilla theory modifies operators as:
\begin{equation}\label{eqsupp:operatorafterprojectionancilla}
    \begin{split}
    P_S f_{i;\alpha}P_S=&f_{i;\alpha},\\
    P_S f^\dagger_{i;\alpha}P_S=&f^\dagger_{i;\alpha},\\
       P_S \psi_{i;\alpha} P_S=&-\sqrt{3}f_{i;\alpha},\\
       P_S \psi^\dagger_{i;\alpha} P_S=&\frac{\sqrt{3}}{3}f^\dagger_{i;\alpha}.\\
    \end{split}
\end{equation}
In the renormalized mean field theory, the Gutzwiller projection $P_G$ modifies the operators as:
\begin{equation}\label{eqsupp:operatorafterprojectionGutzwiller}
    \begin{split}
        P_G \tilde{f}_{i;\alpha} P_G = & f_{i;\alpha},\\
       P_G \tilde{f}^\dagger_{i;\alpha} P_G = & f^\dagger_{i;\alpha},\\
       P_G \tilde{\psi}_{i;\alpha} P_G = & -\sqrt{3}f_{i;\alpha},\\
       P_G \tilde{\psi}^\dagger_{i;\alpha} P_G = & \frac{\sqrt{3}}{3}f^\dagger_{i;\alpha}.
    \end{split}
\end{equation}
We thus conclude that the Gutzwiller projection operator $P_G$ has the same effect as $P_S$ in the ancilla formulation. Accordingly, the $\psi$ fermion in the ancilla theory can be mapped to the $\tilde{\psi}$ fermion in the renormalized mean field theory, preserving the operator correspondence under projection.

\subsection{$\gamma(\mathbf{k})\ne 0$ but $\gamma(\mathbf{k})\ll U$}
In a more physical relevant regime with $\gamma(\mathbf{k})\ne0, \gamma(\mathbf{k})\ll0$, the ground state of $H_\mathrm{ancilla}$ is still a Slater determinant, but $c$ and $f,\psi$ are no more decoupled. We need to rotate the wavefunction Eq.~\ref{eqsupp:ancillawavefunction0} at each momentum $\mathbf{k}$ to be the ground state of $H_\mathrm{ancilla}=(R^{(cf\psi)})^{-1}H_{\mathrm{ancilla},\gamma(\mathbf{k})=0}R^{(cf\psi)}$. Then the Slater determinant after rotation can be written as:
\begin{equation}
    \ket{\mathrm{Slater}[c,f,\psi]}=R^{(cf\psi)}\ket{\mathrm{Slater}[c,f,\psi]}_0,
\end{equation}
where,
\begin{equation}
    \begin{split}R^{(cf\psi)}=&1-\sum_{\mathbf{k},i,\alpha,\beta} \Big(\frac{\gamma_{\alpha\beta}(\mathbf{k}) }{\Phi}\big(\frac{\mu_\psi-\mu}{\Phi}f^\dagger_{i;\alpha}+\psi^\dagger_{i;\alpha}\big) c_{1;\mathbf{k};\beta}\\ 
    &-\mathrm{H.c.}\Big)+o(\frac{\gamma^2(\mathbf{k})}{U^2}).\end{split}
\end{equation}
After obtaining $\ket{\mathrm{Slater}[c,f,\psi]}$, the remaining calculation is similar to what we did in the case with $\gamma(\mathbf{k})=0$. Making use the relation in Eqs.~\ref{eqsupp:operatorafterprojectionancilla},\ref{eqsupp:operatorafterprojectionGutzwiller} and substituting $\Phi=\sqrt{3}U/4,\mu_\psi=U/4+2U(1-\kappa),\mu=0$, we obtain the final ancilla wavefunction as:
\begin{equation}
\begin{split}
    \ket{\Psi_\mathrm{ancilla}}=&P_S R^{(cf\psi)}\ket{\mathrm{Slater}[c,f,\psi]}_0\otimes\ket{\Psi_{\psi'}}\\
    =&\Bigg(1-\sum_{\mathbf{k},i,\alpha,\beta}\Big(\big(\frac{\mu_\psi-\mu}{\Phi}+\frac{\sqrt{3}}{3}\big)\frac{\gamma_{\alpha\beta}}{\Phi}(\mathbf{k})\\
    &f^\dagger_{i;\alpha}c_{1;\mathbf{k};\beta}-\big(\frac{\mu_\psi-\mu}{\Phi}-\sqrt{3}\big)\\
    &\frac{\gamma^*_{\beta\alpha}}{\Phi}(\mathbf{k})c^\dagger_{1;\mathbf{k};\beta}f_{i;\alpha}\Big)\Bigg)\ket{\Psi_\mathrm{ancilla}}_0\\
    =&\Bigg(1-\sum_{\mathbf{k},i,\alpha,\beta}\Big(\frac{4(10-8\kappa)\gamma_{\alpha\beta}(\mathbf{k})}{3U}f^\dagger_{i;\alpha}c_{1;\mathbf{k};\beta}\\
    &+\frac{4(8\kappa-6)\gamma^*_{\beta\alpha}(\mathbf{k})}{3U}c^\dagger_{1;\mathbf{k};\alpha}f_{i;\beta}\Big)\Bigg)\ket{\Psi_\mathrm{ancilla}}_0.
\end{split}
\end{equation}
We can perform a similar procedure to get the Gutziwiller projected wavefunction $\ket{\Psi_\mathrm{GP}}$ at nonzero $\gamma(\mathbf{k})$ while $\gamma(\mathbf{k})\ll U$. Since $g_\gamma\approx1$ while $\gamma(\mathbf{k})\ll U$, the mean field Hamiltonian $H_\mathrm{MF}$ in Eq.~\ref{eqsupp:HSPnewbasis} is the same as Eq.~\ref{eqsupp:ancilla_mf}. Therefore, the ground state of Eq.~\ref{eqsupp:HSPnewbasis} can be written as:
\begin{equation}
    \ket{\Psi_\mathrm{GP}}=R^{(c\tilde{f}\tilde{\psi})}\ket{\Psi_{\mathrm{GP}}}_0,
\end{equation}
where $R^{(c\tilde{f}\tilde{\psi})}$ is $R^{(cf\psi)}$ after replacing $f,\psi$ with $\tilde{f},\tilde{\psi}$. The final Gutziwiller projected wavefunction can be calculated as:
\begin{equation}
    \begin{split}
        \ket{\Psi_\mathrm{GP}}=&P_G R^{(c\tilde{f}\tilde{\psi})}\ket{\mathrm{Slater}[c,\tilde{f},\tilde{\psi}]}_0\\
    =&\Bigg(1-\sum_{\mathbf{k},i,\alpha,\beta}\Big(\frac{4(10-8\kappa)\gamma_{\alpha\beta}(\mathbf{k})}{3U}f^\dagger_{i;\alpha}c_{1;\mathbf{k};\beta}\\
    &+\frac{4(8\kappa-6)\gamma^*_{\beta\alpha}(\mathbf{k})}{3U}c^\dagger_{1;\mathbf{k};\alpha}f_{i;\beta}\Big)\Bigg)\ket{\Psi_\mathrm{GP}}_0,
    \end{split}
\end{equation}
which is equivalent to $\ket{\Psi_\mathrm{ancilla}}$ since $\ket{\Psi_\mathrm{ancilla}}_0=\ket{\Psi_\mathrm{GP}}_0$.

\end{document}